\documentclass[twocolumn,aps,showpacs]{revtex4}
\usepackage{graphicx}
\begin{document}
\setlength{\unitlength}{1mm}
\bibliographystyle{unsrt} 
\title{Berry's Phases for Arbitrary  Spins  Non-Linearly  Coupled  
 to External Fields. \\ Application to  the Entanglement of  $N> 2$ Non-Correlated  One-Half Spins}
 \author{ Marie-Anne Bouchiat}
\affiliation{Laboratoire Kastler-Brossel, CNRS, UPMC, \'Ecole Normale Sup\'erieure,   
24, rue Lhomond, 75005 Paris France,}
\author{ Claude Bouchiat}
\affiliation{Laboratoire de Physique Th\'eorique de l'\'Ecole Normale Sup\'erieure, CNRS, UPMC,  
24, rue Lhomond, 75005 Paris France.}
\date{September 10, 2010}
 \newcommand \be {\begin{equation}}
\newcommand \ee {\end{equation}}
 \newcommand \bea {\begin{eqnarray}}
\newcommand \eea {\end{eqnarray}}
\newcommand \nn \nonumber
\def \(({\left(}
\def \)){\right)}
 \def \va{{\mathbf{a}}}
 \def \vb{{\mathbf{b}}}
\def \vs{{\mathbf{s}}}
 \def  \vS{{\mathbf{S}}}    
 \def \vI{{\mathbf{I}}}
 \def \vr{{\mathbf{r}}}
 \def \vF{{\mathbf{F}}}
 \def \vr{{\mathbf{r}}}
\def \vp{{\mathbf{p}}}
\def \vE{{\mathbf{E}}}
\def \vX{{\mathbf{X  }}}
\def \vB{{\mathbf{B}}}
\def \ve{{\mathbf{e}}}
\def \vk{{\mathbf{k}}}
\def \vC{{\mathbf{C}}}
\def \vD{{\mathbf{D}}}
\def \vd{{\mathbf{d}}}
\def  \vSig{{\mathbf{\Sigma}}}
\def  \veps{{\mathbf{\epsilon}}}
\def  \En{{\mathrm{E}}}
\def\bra{\langle}                                                                                                                                                                                                                                                                    
\def\ket{\rangle}
\def\wt{\widetilde}
\begin{abstract} 
We derive the general formula giving the Berry's Phase for an arbitrary spin, having both  magnetic-dipole  and electric-quadrupole couplings with external  time-dependent fields. We assume that the ``effective" electric and magnetic fields  remain orthogonal during the  quantum cycles. This mild restriction has many advantages. It provides simple symmetries leading to useful selection rules. Furthermore, the  Hamiltonian parameter space coincides  with the density matrix space for a spin $S= 1$. This implies a mathematical identity between Berry's phase and the Aharonov-Anandan phase, which is lost for $S>1$. We have found, indeed, that new physical features of Berry's phases emerge for integer spins $\geq 2$.  
We  provide   explicit numerical results  of Berry's phases for $ S=2,3,4$.  For any spin,
 one finds easily    well-defined regions of the parameter space  where  instantaneous eigenstates remain widely separated. We  have decided  to stay within them and not do deal with non-Abelian Berry's phases. 
We present a thorough and precise analysis of the non-adiabatic corrections with separate treatment for periodic and non-periodic Hamiltonian parameters. In both cases, the accuracy for satisfying the adiabatic approximation within a definite time interval is considerably improved if one chooses for the time derivatives of the  parameters a time-dependence having a Blackman pulse shape. This has the effect of taming the non-adiabatic oscillation corrections which could be generated by a linear ramping. For realistic experimental conditions, the non-adibatic corrections can be kept  below the 0.1{ \%} level.
For a class of quantum cycles,  involving as sole periodic parameter the precession angle of the electric field around the magnetic field, the corrections 
odd upon the reversal of the associated rotation velocity can be cancelled exactly 
if the quadrupole to dipole coupling is chosen appropriately (``magic values").
The even ones  are eliminated by taking  the  difference of the Berry Phases generated
 by  two ``mirror" cycles.
  We end by a possible application  of the theoretical tools developed in the present paper. We propose a way to perform an holonomic entanglement of  $N$  non-correlated 
one-half spins by  performing adiabatic cycles  governed  by a Hamiltonian  given as  
  a non-linear  function of the total spin operator $ \vS$,  defined as  the sum of the $ N $ individual spin  operators. The basic idea behind this proposal is the mathematical
fact that any non-correlated states can be expanded into eigenstates  of ${\vS }^2 $ and 
$ S_z$.The same eigenvalues appear several times in the decomposition  when $N>2$ but all these  states differ by their symmetry properties under the $N$-spin permutations. 
The case $N=4$ and $S_z=1$ is treated explicitly and a maximum  entanglement is achieved.
 
\pacs  {03.65.Vf, 03.67.Bg, 03.75.Dg, 37.25.+k} 
\end{abstract}
 \maketitle
 \tableofcontents
\section{Introduction}
 About twenty five years ago, Geometry made a new entrance in Quantum Mechanics 
 with the discovery of geometric phases \cite{Berry,simon,AAphas}. These  developments hinge upon the
 simple  fact that all the physical information relative to an isolated system described
 by a  pure quantum state $ \vert \Psi  \ket $ is contained in the density matrix 
 $ \rho = \vert \Psi  \ket\bra \Psi \vert$.  This mathematical object has two important
 properties: i) it is invariant upon the abelian gauge transformation  
 $  \vert \Psi  \ket \rightarrow \exp( i \,\chi)\, \vert \Psi  \ket $ where $\chi $ 
 is an arbitrary real number, ii) it satisfies the non-linear relation $  \rho^2= \rho $.
 The density matrix space $ \mathsf{E} ( \rho ) $ has clearly a non-trivial
 topology. A geometric phase is acquired when the quantum system performs a time evolution along a closed  circuit on $ \mathsf{E} ( \rho ) $, and satisfies  at every instant $t$
 the so-called ``parallel transport" condition $  \bra \Psi\vert \frac{d}{dt} \Psi \ket=0 $.
  The above terminology reflects the fact that the set of quantum vector states 
  associated  with a given  density matrix $ \rho $  can be viewed as the ``fibre'' of  a  linear ``fibre bundle"  constructed above the  ``base space'' $ \mathsf{E} ( \rho ) $. 
  
 Our work deals mainly with adiabatic quantum cycles performed 
 within a definite time  interval $ 0\leq t \leq T $.  They are 
  generated by an Hamiltonian depending on a set of parameters $ \vX(t) $, assumed to be a system of coordinates   for a  differential manifold    $ \mathsf{M} $. In this   way,
  an adiabatic quantum cycle generates a mapping of a closed circuit drawn 
  upon  the  parameter space  $ \mathsf{M} $ onto a closed loop    upon the density   matrix space $ \mathsf{E} ( \rho ) $. The Berry phases can then be viewed 
  as the geometric phases associated with  this particular class of quantum cycles.
 
   At this point, a question 
  arises naturally: how can one extract the geometric phase since $ \rho $ is  phase independent?
  The answer lies within the Superposition Principle of Quantum   Mechanics: 
  any linear combination of two quantum states  $\vert \Psi_1 \ket $ and $\vert \Psi_2\ket $   relative to a given quantum system,
    $\vert \Psi_{1\,2}  \ket= c_1  \vert \Psi_1 \ket + c_2 \vert \Psi_2 \ket $ is  an accessible state for  the system.
    In the present context, such a construction will be achieved via the interaction of the system with   specific   classical  radio-frequency fields, using the so-called Ramsey pulses \cite{ram}. 
     The density
      matrix associated with $\vert \Psi_{1\,2}  \ket$ is 
  $ \rho_{1\,2} = \vert \Psi_{1\,2}  \ket \bra \Psi_{1\,2} \vert =\vert c_1  \vert ^2  \rho_1 +\vert c_2  \vert ^2  \rho_2
  + \Delta \rho_{1\,2} $. The crossed contribution: 
  $\Delta \rho_{1\,2}  =  c_1 \, c_2^{*} \vert \Psi_1  \ket\bra \Psi_2 \vert +h.c. $ contains all 
  the information needed to obtain the  difference of the geometric phases acquired by the states 
  $\vert \Psi_1 \ket $ and $\vert \Psi_2\ket $  during an adiabatic quantum cycle.
   In some experimental schemes  considered in a forthcoming paper \cite{bou3}
   the  geometric phase  acquired by  $\vert \Psi_2\ket $  will  be  {\it  a priori} zero, and the measurement will yield directly  the phase acquired by $\vert \Psi_1\ket $.

  A huge amount of work has been triggered by 
  the publication of M. Berry's paper  \cite{Berry} on the phase acquired by a quantum state at
the end of its adiabatic evolution along a closed cycle. Among
 the flurry of following papers, two selections have been presented
  in comprehensive  reviews \cite{wil,Anand} providing guides to the extensive literature 
on the subject. There also exist pedagogical presentations and a thorough review of manifestations 
giving many examples drawn from a great variety of physical systems \cite{SciAm, Holst,Pines}. 
  More recently, Berry's phase has become a topic of renewed interest, both theoretical and 
experimental, with regard to improved information processing and more specifically 
for its potential use in quantum computing \cite{Sjo,jones}. At the same time, its unwanted manifestations 
in fundamental precision measurements, have to be carefully investigated \cite{Comm,Pend,hinds}. Conversely, in other settings, Berry's phase 
might be just the right tool to detect still unobserved  effects, such as parity violation in atomic 
hydrogen \cite{nacht,DeK}.     
 
  This paper presents a detailed study of Berry's phases resulting from adiabatic quantum cycles 
performed by an  arbitrary spin, interacting non-linearly with external electromagnetic fields. We assume that the Hamiltonian, governing the time evolution,  involves  both linear and
quadratic couplings of the  spin operator $\vS$.  The addition of a quadratic spin coupling 
generates new features which, as we shall see, can show up for $S>1$. The first purpose of this paper is to develop the formalism to calculate them, whatever the value of $S$, integer or half-integer. 
In addition, we shall give explicit numerical results for several spin values. 
 Throughout this paper,
we shall deal with the following set of quadratic spin  Hamiltonians:
\be
H(\vB(t),\vE(t))=\gamma_S \; \vS \cdot  \vB(t) + \gamma_Q\;  (\vS 
\cdot \vE(t))^2  \, . 
\label{themodel}
\ee

We have found in the literature only a few papers 
 dealing with the Berrry's phase  for a spin submitted to a time-varying quadrupole 
interaction \cite{tycko},\cite{pines1}.   
These deal with nuclear quadrupole resonance spectra 
in a magnetic resonance experiment involving sample rotation. However,
 the absence of any magnetic interaction is assumed, which leads to  level-degeneracy. 
 These papers generalize the
Berry's phase to the adiabatic transport of degenerate states. In this case, the geometric 
phase is replaced by a unitary matrix given by the Wilson loop integral of a $ SU(n)$ non-abelian   gauge potential  where $n$ is the dimension of the eigenspace associated
 with a given degenerate quantum level \cite{wilzee,kir,zee}. 
  In contrast, in our work we are interested in situations  where the energy levels    
 of the instantaneous Hamiltonian stay widely separated.
 
We impose on the model the extra constraint $\vE \cdot \vB = 0$. 
As shown previously by one of us (C. B.) and G. Gibbons \cite{bou1,bou2}, the parameter space of $H(\vB,\vE)$ with $\vE \cdot \vB = 0$  is  isomorphic to the space   $ \mathsf{E} ( \rho )$
  relative to  spin-one states, namely,the complex projective plane $ \vC P^2 $.
In \cite{bou2}, the   Berry's phase generated by 
the  Hamiltonian  (\ref{themodel}) for  $S=1$,  is   found  to be  mathematically identical to the Aharonov-Anandan (A.A.) phase, if  one uses  an  appropriate parametrization  of $ \vC P^2 $. 
 However, the physical contents are in general  different, 
 since, in contrast to the Berry's phase, the A.A. phase 
 is not restricted to {\it adiabatic} quantum cycles.
  Hereafter, we derive an  
expression for Berry's phase relative to (\ref{themodel}) for spins $S>1$. In this case  
 one loses the  identity with  the   A.A.  phase, which involves now
 closed circuits drawn upon the  larger projective  complex spaces  $\vC P^{2 S}$.
 
It is convenient to introduce the rotation $R(t)$ which brings the $x$ and $z$ axes along the $\vE$ and $\vB$ fields and the associated unitary transformation $ U(R(t))$. This allows us to rewrite the spin Hamiltonian (\ref{themodel}) as $ H(\vE, \vB)= \gamma_S B\; U(R(t)){\cal H}(\lambda )U^{\dagger}(R(t))$, with  the dimensionless Hamiltonian ${\cal H}(\lambda )$ given by ${\cal H}(\lambda )= S_z \hbar^{-1} + \lambda \,S_x^2 \hbar^{-2}$. The  parameter $\lambda = \hbar \gamma_Q E^2/ (\gamma_S B)$, combined with the Euler angles $\theta, \varphi, \alpha$ relative to $R(t)$   leads
to  a convenient set of   coordinates for $\vC P^2 $.
 A quantum cycle within the time interval $0\,,T $ satisfies the boundary conditions $ \theta(T_c) =\theta(0), \,\varphi(Tc)= \varphi(0)+ 2 n_{\varphi} \pi , \,\lambda(T_c)= \lambda(0),\,\alpha(T_c) = \alpha(0) +  n_{\alpha}  \pi $, where $n_{\varphi}$ and $n_{\alpha}$ are arbitrary integers.
 The eigenstates of ${\cal H}(\lambda )$, $\hat \psi (\lambda, m)$ are labeled with a magnetic number $m$ by requiring that the analytical continuation of   $\hat \psi (\lambda, m)$ towards $\lambda=0$ coincides with the angular momentum eigenstates  $\vert S, m\ket$. Thanks to the constraint $\vE\cdot \vB=0$,  ${\cal H}(\lambda )$ has several discrete symmetries we shall use throughout this paper. The ``$m$'' parity $(-1)^{S-m}$, associated with a $ \pi$-rotation around $\hat z$,  is a good quantum number for ${\cal H}(\lambda )$. Therefore ${\cal H}(\lambda )$ can mix states $\hat \psi (\lambda, m_1)$ and $\hat \psi (\lambda, m_2)$ if and only if $\vert m_1 -m_2 \vert $ is an even integer. This selection rule implies that  
$\hat \psi (\lambda, m)$ can be written  as a direct sum of even and odd blocks. This  greatly simplifies   the construction of the eigenstates of $H(\vE, \vB)$, 
 and the Berry-phase determination.

After some manipulations involving Group  Theory techniques, Berry's phase is written as a loop integral in the parameter space $\vC P^2 $:
\be
\beta(m,\lambda)= \oint p(m,\lambda) (\cos{ \theta} d\varphi + d\alpha)  -m (d\varphi + d\alpha)
\ee 
where $\hbar p(m,\lambda)$ is the average value  of $\vS$
along the $\vB $ field direction. This quantity is obtained by taking the gradient of the eigenenergies of  $H(\vE, \vB)$ with respect to $\gamma_S B$. 
We have performed an explicit calculation  of the eigenvalues of  ${\cal H}(\lambda )$ for the spins $S=2,~3$ and 4 for values of $\lambda$ running from 0 to 2, 1.4 and 1.2 respectively (Fig.1). 
A superficial look at the above expression of $\beta(m,\lambda)$
may give the impression that our result is, after all,  not so different from the case of a pure dipole coupling. However, there are specific features associated with the quadrupole coupling which appear more easily for special cycles where $\lambda$
 and $\alpha$  are the only time varying parameters. The particular case $S=2$, $m=0$
is especially instructive to this respect. We recall  that Berry's phase associated with a pure dipole coupling is vanishing  for an arbitrary  cycle with  $m=0$ state.  Performing  
the above simple cycle  with the state $ S=2 ,\, m=0 $  and assuming a constant quadrupole coupling,  $\lambda(t)=1$,  one  obtains from  a look at  curves of Fig. 1 the quite 
remarkable result $\beta(m=0,\lambda)= \oint p(0,\lambda) d\alpha = \pi \; (mod \;2\pi).$

Our motivation for this paper is to make contact with experiment, having in mind the spectacular progress of atomic interferometry. One has then  to face the problem of the practical realization of a quadrupole coupling, having a magnitude comparable to the magnetic dipole one. For alkali atoms 
this is  unrealistic if the Stark shift arises from a static E-field.  The ac Stark shift \cite{cohen} induced by a light beam  is much more flexible. However, in most cases,  reaching   values  of $\lambda$ close to 1    requires the  tuning of the beam frequency to be  so close to an  atomic line that  
the  ac Stark effect  induces an instability of the ``dressed"  atomic ground state. In typical cases this implies  stringent constraints upon the duration of the Berry's cycle. As a result, the question of the validity of the adiabatic approximation becomes crucial. This has led us to devote a full section of this paper to the precise evaluation of the non-adiabatic corrections. Our theoretical analysis is illustrated by numerical results for a few relevant cases.  It provides guide lines for our forthcoming paper, devoted to experimental proposals. 

 Our analysis of the non-adiabatic corrections  proceeds in two steps. We first
 deal with the corrections associated with the time derivatives  of the Euler angles $\dot{\theta}, \dot\varphi, \dot{\alpha}$.  A convenient approach  is the study of the Berry's 
 cycle in the rotating frame attached to the  $\vB$, and $\vE $ fields, by performing upon  the laboratory
 quantum state $ \Phi(t) $ the unitary transformation $\wt  \Phi(t)=  U^{-1}(R(t)) \Phi(t) $. The corresponding Hamiltonian is obtained by adding to  
 $\gamma_S B {\cal H}(\lambda )$ the extra term $  \gamma_S \vS \cdot \Delta \vB (t) $,
 where $\Delta \vB (t)$ is the magnetic field generated by the Coriolis effect.  The 
 longitudinal component ${\Delta \vB }_{//}   =    - \gamma_S ^{-1} \, ( \cos\theta\; \dot 
\varphi+ \dot \alpha ) \,\hat z$, gives rise to a pure dynamical phase shift at the end of the Berry's cycle. As expected it incorporates $ \beta (m) -  m \oint (d\varphi + d\alpha) $ 
 at its lowest  order contribution with respect to  $\eta=-{\Delta B }_{//}( \gamma_S B)$.
 When $ \alpha $ is the sole varying Euler parameter, the higher order  terms in $\eta $ 
 gives the complete set of non-adiabatic corrections. In addition, we show that the odd-order ones $ \propto \eta^{2 n+1} $ vanish exactly if 
 one chooses for $\lambda $ the ``magic" value $\lambda^* (\eta)$. On the other hand the transverse 
 component  $ {\Delta \vB}_{\perp} $, proportional to $  \sin\theta\; \dot 
\varphi $, presents risks: it involves a linear combination of the 
spin operators $S_x $ and $S_y$ and induces a mixing with opposite $m$-parity states  
( $ \Delta m =\pm 1 $),  possibly nearly degenerate unless stringent constraints
are imposed upon $\lambda $.
As regards the non-adiabatic  corrections associated with $\dot \lambda $, we concentrate our attention to  the ramping process  of 
 $ \lambda(t)$ from $\lambda(0)=0 $ to $ \lambda(T) \sim 1$,  the Euler  angles 
 keeping fixed values.  Our method can be viewed as an extension  of the rotating frame approach.  We introduce the  unitary transformation which makes ${\cal H}(\lambda )$  diagonal within the $ \vert S \, m \ket $ basis. The time evolution  equation  acquires an extra
non-diagonal matrix  $ \Delta {\cal H}(t) \propto \dot{\lambda}(t)$ which has the same effect as an rf  pulse with sharp edges if  $ \lambda(t)$
increases linearly with $t$.    
This leads to  rather large oscillating non-adiabatic
corrections exhibited in our work (see Fig.\ref{adiabla}). 
  The  standard procedure to tame them
  out is to give to $\dot \lambda(t)$ a Blackman pulse shape \cite{Black}. 
   
   In the last section, we propose  an ``holonomic" procedure  for the 
  entanglement of $N$ non-correlated 1/2-spins (or N Qbits.)
  The  basic tools are Berry's cycles   generated   
  by a Hamiltonian,  formally identical to $H(\vB(t),\vE(t)) $, 
 except that, now, $ \vS $ is meant to be {\it the total spin operator}
of the $N$ spins, $\sum_{i=1}^{N} \vs_i .$ 
  The method is based on a known mathematical property: 
any non-correlated  $N$ 1/2-spin state can be expanded into a sum of 
 $\vS^2$ and $ \vS_z$ eigenstates. A given eigenvalue of $ \vS^2$  will
appear several times if $ N> 2$, but all the angular momentum 
states  $ \Psi_{ S\, M}^i$  have different  symmetry properties 
under permutations of the $N$ spins, which leave $H(\vB(t),\vE(t)) $ invariant by construction. Thus  
the states $ \Psi_{ S\, M}^i$, organized into an orthogonal basis, behave as if they were associated with isolated spins.   
 An initial non-correlated state, written as 
the sum $ \sum _{i\,, S} a_{S\,M}^i \Psi_{S\, M}^i $, is transformed at 
the end of Berry's cycle  into  $ \sum_{i ,\, S}  a_{S\,M}^i \exp{ i \beta (S \, M)}\Psi_{S,\, M}^i $. With  an appropriate  choice of the cycle, we have been able 
to achieve maximum entanglement 
for $N=4\,,\, S_z =1$.
      
   \section{The instantaneous  eigenfunctions of  $H(\vB,\vE)$ for a given adiabatic cycle}
 In this section we construct the  instantaneous  eigenfunctions of
$H(\vB,\vE)$ for an arbitrary adiabatic cycle. The result is put under a form 
   well adapted to the calculation  of Berry's phase by group theoretical methods. 
 Our method applies to both integer and half-integer spins.
\subsection{Instantaneous spin Hamitonian.  Symmetry properties}
 As a preliminary   step,  it is convenient to   study  the  particular  field  configuration  where $\vE$ and $\vB $ are along the $x$- and $z$-axes respectively, 
 $\hat n= \hat x,~\hat b= \hat z$, and write: 
\be
\hat H(B,E)= \gamma_S B \; S_z + \gamma_Q E^2 S_x^2 \,, 
\ee
where the term $ \vS^2/3= \hbar^2 S(S+1)/3$, that plays no role in the calculation
of Berry's phase has been omitted.  In other words,  $\hat H(B,E)$  is the spin Hamiltonian  in the  frame 
attached  to  the fields $\vB(t) $ and $\vE(t) $, ignoring their time-dependence. 
For the  explicit calculations to be performed in the  cases $S>1$, it is convenient 
to  introduce the dimensionless Hamiltonian ${\cal H}(\lambda) $:
\bea
 \widehat H(B,E)&=&\gamma_S B \hbar\; {\cal H}(\lambda) ;\nonumber \\
  {\cal H}(\lambda) &=&  S_z/\hbar + \lambda\, ( S_x/\hbar)^2 ;\nonumber \\
  \lambda& = &\hbar \gamma_Q E^2/ (\gamma_S B ).
 \eea
  
It is important to note that  ${\cal H}(\lambda)$ is invariant under three  
transformations. The first,  ${\cal T}_1$,  corresponds to the reflexion with
respect to the
$x,y$ plane, which changes $S_x, S_y $ into $-S_x$ and $-S_y$, but has no effect on $S_z$  
\be
 {\cal T}_1^{-1} \, S_x \, {\cal T}_1 = -S_x ,\hspace{3mm} {\cal T}_1^{-1} 
\, S_y \, {\cal T}_1 = -S_y,
\hspace{3mm} {\cal T}_1^{-1} \, S_z \, {\cal T}_1 = S_z . 
\label{Sym1}
\ee
 The second transformation, ${\cal T}_2$, is the product of the reflexion with respect to the $y,z$
plane by the time reversal operation. The transformation of the spin operator under ${\cal T}_2$
obeys the relations: 
\be
 {\cal T}_2^{-1} \, S_x \, {\cal T}_2 = -S_x,\hspace{3mm} {\cal T}_2^{-1} 
\, S_y \, {\cal T}_2 =
S_y ,\hspace{3mm} {\cal T}_2^{-1} \, S_z \, {\cal T}_2 = S_z  .  
\label{Sym2}
\ee
The third transformation is a rotation of $\pi $ around the $z$ axis, which 
changes $S_x$ and $S_y$ into $-S_x$ and $-S_y$ respectively, while leaving  $S_z$ unaltered.
\be
 {\cal T}_3^{-1} \, S_x \, {\cal T}_3 = -S_x ,\hspace{3mm} {\cal T}_3^{-1} 
\, S_y \, {\cal T}_3 = -S_y,
\hspace{3mm} {\cal T}_3^{-1} \, S_z \, {\cal T}_3 = S_z .  
\label{Sym3}
\ee
(Note that ${\cal T}_1$ and ${\cal T}_3$   have the same effect on pseudo-vectors 
but opposite effects on vectors.)

 Let us now discuss  some  consequences of these invariance properties. 
 To this end, let us introduce  the eigenvectors $\hat{\psi}(m,\lambda)$ of the Hamiltonian 
${\cal H}(\lambda)$, together with their eigenenergies $   {\cal E}(m,\lambda) $:
\be
{\cal H}(\lambda) \hat {\psi }(m,\lambda)    ={\cal E}(m,\lambda)  \hat {\psi}(m,\lambda). 
\label{calH}
\ee 
All along this work the eigenenergies are supposed non-degenerate. 
 In the limit $\lambda \rightarrow 0$,  the states
$\hat {\psi}(m, \lambda )$  have to coincide with  the angular momentum states $\vert S,m\ket$:
$\vS^2 \vert S,m\ket =\hbar^2 S(S+1) \vert S,m\ket$ and $S_z\vert S,m\ket= \hbar m \vert
S,m\ket$. 
The relations (\ref{Sym1}) imply that the quantum average of $\vSig= \vS/\hbar $ relative to $\vert
\hat\psi(m,\lambda\not=0)\ket$, {\it i.e.} the polarization of the quantum state $\vp(m,\lambda)$,  
lies along the z-axis:
\be
\bra \hat{\psi}(m,\lambda ) \vert \vSig \vert \hat{\psi}(m,\lambda) \ket =  \lbrace 0, 0,  \, 
p(m,\lambda)\rbrace.
\ee
The invariance under ${\cal T}_1$ and ${\cal T}_2$   requires that the off-diagonal elements  of
the alignement tensor vanish: 
$$A_{ij} = \frac{1}{2} \bra  \hat{\psi}(m,\lambda )\vert \{\Sigma_i,\Sigma_j\}Ê
\vert  \hat{\psi}(m,\lambda )\ket =
\delta_{i,j} A_{ii}\; . $$
 
A third important consequence of the above  invariance properties 
 is that  the eigenvectors $\hat{\psi(}m,\lambda )$ 
may be taken as real. This can be verified by
noting that the matrix associated  with ${\cal H}(\lambda)$ in 
the angular momentum basis $\vert S,m\ket $,
using the standard phase convention, is real and symmetric, so that its eigenvectors  are
real. 

 Equation (\ref{Sym3}) implies that the 
 operator associated  with the rotation of $\pi$  around $\hat z$, $ {\cal T}_3=  \exp( -i \pi \, S_z  )$,   commutes with the Hamiltonian $ {\cal H}(\lambda) $. For convenience, let us  introduce  the ``m-Parity ", 
  $ {\cal P}_3= \exp( i S) \,  {\cal T}_3= \exp (i \,\pi ( S-S_z)) $.  The  m-Parity  
  of the angular momentum eigenstate  state  $\vert  S, m\ket$ is  $( -1) ^{(S-m)}$.
Within  the angular momentum basis, the Hamiltonian ${\cal H}(\lambda)$ can then be written as  the  direct sum   of two matrices  acting respectively on the states even and odd with respect to the operator $ {\cal P}_3$:
\be 
 {\cal H}(\lambda)=   {\cal H}_{even}(\lambda)  \oplus  {\cal H}_{odd}(\lambda). 
\ee
\indent As a conclusion, we would  like to stress that the  field orthogonality condition plays an essential role in making the mathematical problem tractable for spins $S >1$.
Otherwise, the problem would  become rapidly  complicated and  any  insight into  Berry's phase physics gets blurred by the algebra.     

\subsection{The   instantaneous  eigenfunctions   of   $H(\vB(t),\vE(t))$ }
 Since we are going to use group theory arguments, it is appropriate
 to recall  some basic facts about the rotation  
group  in  Quantum Mechanics.
 One introduces the unitary operator  $ÊU({\cal R}(\hat u ,\chi)) $  associated with 
 the rotation ${\cal R}(\hat u ,\chi) $  acting on the spin state vectors: 
\be
U({\cal R}(\hat u ,\chi))=\exp\((-i \frac {\hat u \cdot\vS}{\hbar} \chi \))\, , 
\label{GR1} 
\ee 
where ${\cal R}(\hat u,\chi)$ stands for  the rotation $R$ around the unit vector
$\hat u$ by an angle $\chi$ . This operator provides a unitary  representation of   the rotation  group in the sense that it
obeys the multiplication rule:
 $  U(R_1)\, U(R_2) = U( R_1\cdot R_2)$, together with the unitarity relation $ U^{-1}(R)  = U^{\dag}( R)$.
 Applying the above rule  to the case where  $ R_2$ is an infinitesimal rotation, one derives 
  the important  relation:
  \be
U^{-1}({\cal R}(\hat u,\chi))\;\;\vS \; \;U({\cal R(}\hat u,\chi)= {\cal R}(\hat u, \chi) \cdot\vS, 
\label{GR2}
\ee
 which expresses the fact that  the unitary transformation   $ ÊU({\cal R}(\hat u ,\chi)) $ 
  rotates the spin observables.

 Let us associate  with the orthogonal vectors $\vE$ and $\vB$ the
trihedron $(\hat e, \hat b\wedge \hat e, \hat b)$ which can be constructed by applying the  
rotation $R_E(\theta, \varphi, \alpha)$ to the fixed coordinate
trihedron $(\hat x, \hat y, \hat z) $, 
with $\theta, \varphi, \alpha $ denoting the usual Euler angles:
\be
R_E(\theta, \varphi, \alpha)={\cal R}(\hat z,\varphi){\cal R}(\hat y,\theta){\cal R}(\hat z,\alpha).
\label{REu}
\ee          
 To ensure the validity of the adiabatic approximation for the quantum cycles generated by
$H(\vB(t),\vE(t))$, we shall  assume that the Euler angles are {\it  slowly } varying functions of time.
 More  precisely, we shall require  that their time derivatives 
  $ \dot{\alpha}, \dot{\theta}, \dot{\varphi}$, - together  with the time derivative of the Stark-Zeeman coupling ratio
 $\dot{\lambda} $ -  are much smaller than the rate  
 $ \Delta\, E_{min}/\hbar $, where $\Delta\, E_{min}$  stands  
for  the minimum   distance between the energy levels   of    $ \hat H(B(t),E(t)) $.
The adiabatic cycle within the time interval $0  \leq t \leq T$ is specified by the
 boundary conditions involving the two finite  integers $n_{\varphi}$ and $n_
{\alpha}$ :
  \be
\varphi(T)= \varphi(0)+2 n_{\varphi} \, \pi   \, ; \;\ \alpha(T)=\alpha(0)+ n_
{\alpha}\, \pi. \nonumber\\
 \label{boundcond}
 \ee
 Note that in  contrast to the  periodic variables $ \varphi(t)$  and $\alpha(t) $,
   $\lambda(t)  $    and $\theta(t) $  recover their initial values 
   at the end of the cycle. 
   
   To proceed it is convenient to introduce   the unitary operator $ U(R(t))$    associated with 
 the rotation $R(t)$,  given by:
       \be
     R(t)= {\cal R}(\hat z,\varphi(t)){\cal R}(\hat y,\theta(t)){\cal R}(\hat z,\alpha(t)).
      \label{RotE}
      \ee 
Since    $ U(R(t))$  belongs to   a unitary representation of the rotation group, it can be written,
  using equation (\ref{RotE}), as the following operator product:
  \be
U(R(t))=\exp{-i\frac {S_z}{\hbar} \varphi } \, \cdot \, \exp{-i\frac {S_y}{\hbar} \theta}  \,
\cdot \,   
\exp{-i\frac {S_z}{\hbar} \alpha} \,. 
\label{UR}
\ee

   Using  the relation (\ref{GR2}) with $ {\cal R(}\hat u,\chi)= R(t)^{-1} $ and the identity 
  $ R(t)^{-1} \va \cdot \vb =\va \cdot R(t) \vb $,  it is straightforward to derive the important relation:
  \be
  U(R(t)) {\widehat H}(B(t),E(t)) U^{\dag}( R)=    
 H(\vB(t),\vE(t))
  \ee 
 As a consequence,   the   wave functions $ \Psi(m,t)$   defined as 
\be
\Psi(m,t)= U(R(t)) \hat \psi(m,\lambda(t)), 
\label{instwf}
\ee
are  instantaneous  eigenfunctions of  $  H(\vB(t),\vE(t)) $ with eigenvalues 
$\En(m,B(t),E(t))$ :
\bea
 H(\vB(t),\vE(t))\, \Psi(m,t) \nonumber &=&   U(R(t)) {\widehat H}(B(t),E(t))\, \hat \psi(m,\lambda(t) )    \nonumber \\
 &=& \En(m,B(t),E(t))\Psi(m,t) \\ 
   \En(m,B(t),E(t) )&=& \gamma_{S} \hbar  B(t)  {\cal E}(m,\lambda(t)) .
 \eea
 We would like to stress  that the instantaneous wave functions   $ \Psi(m,t)$   have, by construction,
  a well defined phase since, as shown  previously, the state vectors $\hat \psi(m,\lambda(t)) $ can all be taken as real.

     \section{Explicit calculation  of the eigenfunction parameters 
     for Berry's phase evaluation}
       \begin{figure*}
 \centering\includegraphics[  ]{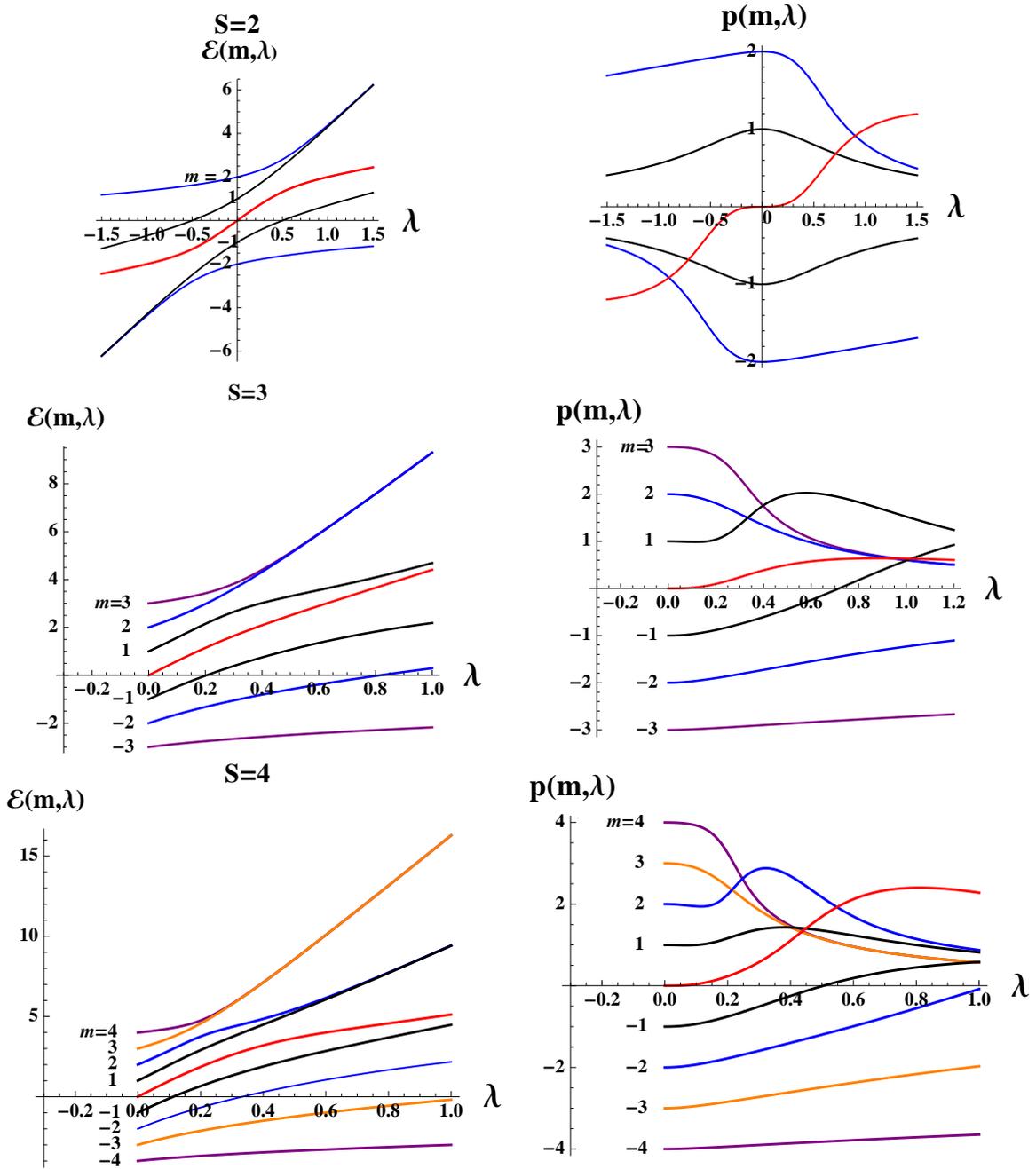}
\caption{ \small Reduced  energies ${\cal E}(m,\lambda) $  and polarizations $ p(m,\lambda) $ 
versus $\lambda$ for $ S=2, 3$ and 4. {\bf Since} ${\cal E}(m,\lambda=0) =p(m,\lambda) =m$, 
{\bf intersection of the curves with the vertical axis  $ \lambda=0$ indicate their $m$ labeling}.  
The $S=2$ curves  exibit  the symmetry relations 
$ {\cal E}(m,\lambda)=-{\cal E}(-m,-\lambda) $ and $  p(m,\lambda)=  - p(-m,-\lambda) $, derived in the text. For $m>0$ there is a clear evidence for the  degeneracies
 expected  when  $ \lambda \gg 1 $. }
 \label{fig1}
\end{figure*}
 We  present for $S=2, 3$ and 4, explicit calcutations  of the 
 eigenenergies  ${\cal E}(m,\lambda)$  and the polarization of the eigenstates,
 which, for symmetry reason, is  along the $ \vB$ direction 
  $p(m,\lambda) =\bra \hat{\psi}(m,\lambda)\vert S_z \vert \hat{\psi}(m,\lambda)\ket $. (In this section, for the  
sake of simplicity we use  a unit system where $ \hbar =1$). 
   We construct 
the matrix associated with  ${\cal H}(S,\lambda)$within the angular momentum
basis:$ \vert S\, m \ket$.
It is convenient to rewrite the Hamiltonian in terms of the operators $S_z,  S^
{+} = S_x+ iS_y$
and $S^{-} = S_x - iS_y$ 
\bea
{\cal H}(S,\lambda) &= &S_z + \lambda S_x^2 \; ,  \nonumber \\
&\equiv& S_z - \frac{\lambda}{2} 
S_z^2 +
\frac{\lambda}{4}( (S^{+})^2 + (S^{-})^2 ) + \frac{\lambda}{2} S(S+1)\,. 
 \nonumber 
\eea
Using textbook formulae, for the matrix elements  
of   $(S^{\pm})^2 $  in the $ \vert S\, m \ket $ basis it is easy to write 
${\cal H}(S,\lambda)$ in matrix form.
   Using the 
invariance under the symmetry $ {\cal T }_3$  introduced in subsection II.A,  
     $ {\cal H}( S,\lambda) $ can be written as
 a direct sum  of two matrices acting respectively upon the states  $ \vert S m \ket$
with  even and odd values of $S-m$,
$ {\cal H}( S,\lambda) =    {\cal H}_{even}( S,\lambda)   \oplus  
{\cal H}_{odd}( S,\lambda), $ one of order $S$ and the other of order $(S+1)$ 
depending on the parity of S.
     For $ S=1$ and 2   one finds easily:
       \bea 
       {\cal H}_{odd}( 1,\lambda)&=&  \left(
\begin{array}{cc}
    \lambda / 2+1 &   \lambda /2 \\
    \lambda /2 & \lambda /2 -1
\end{array}
\right) , \nonumber \\
{\cal H}_{even}( 2,\lambda) &=&\left(
\begin{array}{ccc}
 \lambda +2 &  \sqrt{3/2}\;\lambda  & 0 \\
 \sqrt{3/2}\;\lambda  & 3 \lambda  &  \sqrt{3/2} \;\lambda  \\
 0 &  \sqrt{3/2}\;\lambda  & \lambda -2
\end{array}
\right) , \nonumber\\  
{\cal H}_{odd}( 2,\lambda)&=&  \left(
\begin{array}{cc}
   5 \lambda / 2+1 &  3 \lambda /2 \\
   3 \lambda /2 & 5 \lambda /2 -1
\end{array}
\right) . 
\label{H12}
\eea
The Hamiltonians   ${\cal H}_{even}( 3,\lambda)$  and
 ${\cal H}_{odd}( 3,\lambda)$ as well as ${\cal H}_{even}( 4,\lambda)$  and
 ${\cal H}_{odd}( 4,\lambda)$ are given in Appendix B.
 
The eigenvalues are obtained in a standard way by solving  the two polynomial 
equations: 
\bea 
\hspace{-10mm}\det(x \;\mathbf{l}-{\cal H}_{even}( 2,\lambda))&\equiv&x^3-5 x^2 \lambda +4 x \lambda 
^2-4 x+12 \lambda \nonumber\\
&=&0,  \nonumber\\
 \det(x \;\mathbf{l}-{\cal H}_{odd}( 2,\lambda)) &\equiv& x^2-5 x \lambda +4 \lambda 
^2-1=0.
 \eea       
 The eigenvalue equations  for $ S= 3 $ and $S=4$  can be found 
 in the appendix. 
 Mathematica  codes   yield explicit expressions for the eigenenergies 
 $ {\cal E}(m,\lambda )$.  Although the formulae are  rather  complex, 
 they lead to very accurate  numerical   results for all the values of $ \lambda $ of interest.
  The accuracy is better  than $ 10^{-12}$. This result has been checked using 
numerical Mathematica codes, obtained directly from  the matrix expression.
To calculate  the polarization $p(m, \lambda)$ we have used 
   the fact that it is given by the derivative of the  eigenenergy with 
respect to
$\gamma_S B$ (Hellmann-Feynman theorem). Writing the  eigenvalue of  $\hat H(B,E)$   as:
$  \En(m, B, E )= \gamma_S B  \, {\cal E}(m,\lambda)$, with 
$\lambda= \frac{\hbar \gamma_Q E^2}{\gamma_S B }$, one gets immediately: 
\be
 p(m,\lambda) = 
\frac  {\partial \, \En(m, B, E) }  { \hbar\gamma_S\,\partial\,  B    }=
   {\cal E}(m,\lambda) - \lambda \frac {\partial { \cal E}(m,\lambda)}{\partial 
\,\lambda}.
\label{polarvsla}
\ee
 The results  are given in Fig.{\ref{fig1}  for both the energies and the polarizations. 
   The eigenenergies $ {\cal E}(m,\lambda )$  can be labelled  
 unambiguously since 
 for small values of $\lambda$ the eigenvalues  are equal to  $m$,  up to corrections of 
the order of $ \lambda^2$. 
   In Fig.{\ref{fig1} there is   evidence for  
 near-degeneracies of opposite-parity level  pairs:
  $  {\cal E}(m, \lambda)$ is approaching ${\cal E}(m-1, \lambda) $
    when  $\lambda \gg 1$ for $m=S-q $,  where  $q $ is an  even integer 
    such that $ 0 \leq q <  S $.
   This was expected since in this limit $  {\cal  H}(\lambda) $ is dominated  by the term 
$ \propto  S_x^2 $. The convergence is much  slower for $m<0$ since the positive term
 $ \lambda \,S_x ^2 $ term  has then  to fight against a negative Zeeman effect. 
   There are $ S $ degenerate doublets  with ${\cal E}(m, \lambda)/\lambda \simeq m^2$, 
 as  expected in  the  limit $ \lambda  \gg 1$.
  In addition, by looking at the even-odd (or odd-even) pairs, 
   one sees  clearly that the pair  $ {\cal E}(S, \lambda), {\cal E}(S-1, \lambda)$
   converges, {\it without  crossing},  to the degenerate doublet  
   $ {\cal E}(S, \lambda)\simeq  {\cal E}(S-1, \lambda ) \simeq \lambda    S^2 .$
 The next lower pair ends as the degenerate doublet having
  the energy $ \simeq  \lambda \, (S-1)^2 $ and so on  until one reaches the  isolated level  $ m= - S$, 
 which has no possibility  other than converging
  towards the non-degenerate level with  energy $\lambda \times 0$. 
  While this behaviour is clearly exhibited in the simple case $ S=2 $,   
 only its two first  steps are clearly apparent for $ S=3 $ and 4 , but  we have verified that 
    the above picture  is valid for all values of $m$ by computing the ratios
  $  {\cal E}(m, \lambda)/\lambda $ when  $\lambda \gg 1 $.
 
 A striking feature of the case $S=2$ in  Fig.\ref{fig1} is the symmetry of the plotted curves under the 
transformation $ m  \rightarrow -m \, ,\,\lambda\rightarrow -\lambda $, involving 
 both the eigenenergies and the polarizations:  
 $ { \cal E}(m,\lambda) \rightarrow - { \cal E}(-m,-\lambda)$, 
   $ p(m,\lambda) \rightarrow  -p(-m,-\lambda). $
To prove these  symmetry properties in the general case, it is convenient to perform upon the spin system
a rotation of angle $ \pi $  around the $x$ axis, $ R(\hat x , \pi )$. 
By introducing  the   spin unitary  operator $ U (R(\hat x , \pi ))= \exp (- i \,S_x \, \pi ), $ 
associated  with the rotation $ R(\hat x , \pi )$, one can write 
the transformation law for the Hamiltonian ${\cal  H}(\lambda)$:
$$ U (R(\hat x , \pi )){\cal  H}(\lambda) U^{-1}(R(\hat x , \pi ))= - S_z +\lambda 
S_x^2=
-{\cal  H}(-\lambda). $$
Applying $ U (R(\hat x , \pi ))$ to both sides of the eigenvalue equation  
$  {\cal  H}(\lambda) \hat \psi(m,\lambda)={\cal E}(m,\lambda) \hat \psi(m,\lambda)   
$, one can write:
\bea
U (R(\hat x , \pi )){\cal  H}(\lambda) \hat \psi(m,\lambda) 
&=& -{\cal  H}(-\lambda)  U (R(\hat x , \pi )) \hat \psi(m,\lambda) \nonumber\\
&&\hspace{-10mm}={ \cal E}(m,\lambda)  U (R(\hat x , \pi )  \hat \psi(m,\lambda).
\eea
The state   $U (R(\hat x , \pi )  \hat \psi(m,\lambda)$ is an eigenstate of 
${\cal  H}(-\lambda) $  with the eigenenergy $ -{\cal E}(m,\lambda) $.
It coincides, up to  a phase factor, with the eigenstate  $\hat \psi(-m,-\lambda)$,  
since the effect of  the rotation $ R(\hat x , \pi )$ is to flip the spin component along the $z
$ axis. 
This completes the proof that:
 $$ { \cal E}(m,\lambda) = - { \cal E}(-m,-\lambda) .$$ 
 with the similar relation for $ p(m,\lambda) $ as a consequence of
Eq. (\ref{polarvsla}). 

In addition, the  quantum-averaged polarizations $ p(m,\lambda) $ satisfy the sum rule
\be
\sum_m p(m,\lambda)=0 ,
\ee
 valid for any value of S and $\lambda$. It is readily derived using the fact that $ p(m,\lambda) $ can be evaluated as the partial derivative of the energy with respect to $\gamma_S B$ (Eq. (\ref{polarvsla})) and noting the general structure of the polynomial equation whose solutions provide the eigenenergies. 

 To end this section we would like to say a few words concerning the state $ m=0 $
 which is of particular interest for integer spins $ S \geq 2 $. The fact  that a
  second-order Stark effect can induce a polarization  in  an initially unpolarized 
 state was somewhat unexpected. To gain greater insight we have performed
a perturbation computation in powers of $ \lambda $. Although one has to go to third order to find a non-zero  effect the final result is  given by the rather elegant  formula:
 \be 
   p(0,\lambda )=\frac{1}{8} {\lambda}^3 S (S+2) \left(S^2-1\right) 
    \(( 1+ {\cal O} ( \lambda^2)  \)) .
 \ee            
It  gives a  rather accurate  result for $S= 2$ if $ \vert \lambda \vert \leq 0.4 $ but 
the domain of validity gets smaller for S=3: $ \vert \lambda \vert \leq 0.12 $. Finally we 
would like to point out the  smooth behaviour  of $ p(0,\lambda$) for $ S=2 $ in the 
vicinity of $ \lambda \simeq 1$ and the remarkably simple exact values taken  by  the 
reduced energy and the polarization  at $ \lambda=1$, namely   
${ \cal E}(0,1) =2$ and $ p(0,1)=1$. At the time of writing, it is unclear whether  
this  is a mere numerical accident or an indication of something  more profound.
 
\section{Berry's phase for  adiabatic cycles  generated  by     
 $  H(\vB(t),\vE(t))$ acting on arbitrary spins } 
 We start this section by a mini-review introducing the basic physical and mathematical  
features of Berry's phase  concept. In particular we  derive  the general
 formula giving the Berry' phase in terms of the instantaneous eigenfunctions
 of the time-dependent Hamiltonian generating the adiabatic  quantum cycles.
 Introducing in this formula,  the results of the previous section and 
 relying upon Group Theory arguments, we perform an explicit
 construction of Berry's phase relative to $  H(\vB(t),\vE(t))$, valid for arbitrary spins.
 The final  result is expressed
as a loop integral in the   $  \vC P^2$
 space,   using as coordinates  the Euler angles and the 
parameter $ \lambda $.  
We show that for the case $ S =2$  a circular loop  drawn upon a spherical 
subspace of  $  \vC P^2$  lead to a loop integral very different 
from that of the case $S=1$ which involves a magnetic monopole 
Bohm-Aharonov phase.   
 \subsection{ The  Berry's phase as a physical observable and  topological 
concept: a mini-review}
In this  introductory  subsection, 
 we are going to  follow, in several places, a presentation due to the late Leonard  Schiff, in tribute of its memory. He gave  in  few pages of its  venerable textbook \cite{Schiff}  a correct  and precise treatment of the adiabatic approximation, involving  a non-integrable phase.
To study the adiabatic quantum  cycles generated  by   the 
  Schr\"odinger equation     
     $$  i\,\hbar \frac{\partial}{\partial t} 
   \Phi(t)= H(t)  \, \Phi(t) \hspace{5mm} \text{ where}\hspace{5mm} 
   \,  H(t)=H(\vB(t),\vE(t) ) , $$
it is convenient  to  expand  the solution $\Phi(t)$  in terms of the 
eigenstates of  $H(t)$  :
\bea 
&&\Phi(t) =\sum_{n}  a_{n}(t)   \exp \(( \frac{i}{\hbar} \,\chi(n,t)  \))\Psi(n,t) , 
\label{Phi}   \\
 &&\hspace{-10mm}{\rm where} \;\;\; \chi(n,t)  =  \gamma(n, t) -\int _0^t d t_1  \En(n,B(t_1), E(t_1)) .     \label{expPhi}
 \eea
   The first term   $\gamma(n, t)$ in (\ref {expPhi}) is a phase that vanishes for
    $ t=0 $, but is otherwise arbitrary.
   Its value will be determined by  the reasoning leading to the adiabatic 
approximation. 
   The second term,
   \be  
  \Phi_D(n) = - \frac{1}{\hbar}\int _0^t d t_1  \En(n,B(t_1), E(t_1)),\label{Dynphase} 
  \ee 
 known as the ``dynamical phase",  produces  a  contribution  which  
    cancels  $ H(t) \Phi(t) $ in  the wave  equation. 
  We shall take as the initial condition : $ \Phi(0)=  \Psi(m,0) $  or, in other words,  
$ a_n(0) = \delta_{n,m} $. 
   In this case we can replace the exact Schr\"odinger equation
   by the system of differential  equations involving the expansion coefficients $ a_{n}(t)$:  
   \bea
  &&i \,\hbar \,\dot{a}_n(t) =  a_n(t)\,(\dot{\gamma} (n,t)-
   \bra \Psi(n,t)\vert  i\,\hbar\frac {\partial}{\partial t} \Psi(n,t)\ket) -
  \nonumber \\
   &&\hspace{-8mm}\sum_{k \neq n} \hspace{-1mm}a_{k}(t)   
   \exp {  i (\chi(k,t) -\chi(n,t) ) }  
   \bra\Psi(n,t)\vert i\,\hbar\frac {\partial}{\partial t} \Psi(k,t)\ket . 
   \label{eqSchi}
    \eea
      Since the adiabatic condition requires that $ a_n(t) \approx 1$ whatever $t$, 
    a necessary condition to ensure its validity is to
    cancel the coefficient of $   a_n(t) $  in  the r.h.s of  equation  (\ref{eqSchi}).
       This is  achieved if we make the following choice for the 
``gauge "  $\gamma (n,t) $:
     \be
     \gamma(n, t )  =
   \int_0^t   dt_1\bra\Psi(n,t_1)\vert  \, i\,\hbar\frac {\partial}{\partial t} 
\Psi(n,t_1)\ket .
   \label{Shifgau}
      \ee
   Later on, we shall see that  this  non integrable  gauge   is a basic ingredient in the  
 mathematical expression  of Berry's phase  in terms 
 of the instantaneous wave functions $ \Psi(m,t) $. 
       The next step  to validate  the adiabatic approximation  is to find 
         appropriate conditions allowing
           the  sum  $ {\sum_{n\neq m}\vert a_n(t) \vert }^2  $
         to remain below  a predefined  level  for $ t >0$. This task will be performed in details in Sec. V but for the moment, let us assume that it is achieved.  
                      {\it Within the adiabatic approximation,} the  solution of the Schr\"odinger equation   
       $ i\,\hbar \frac{\partial}{\partial t}  \Phi(t)= H(t)  \, \Phi(t)$, with the 
initial condition $ \Phi(0) = \Psi(m,0) $, is  then  given  by:
       \bea 
      && \Phi_{ADB}(m,t)= \nonumber \\
  &&\hspace{-10mm} \exp \frac{i}{\hbar} \((\gamma(m, t)- \int _0^t \hspace{-1mm} dt_1  
  \En(m,B(t_1),E(t_1)) \hspace{-0.5mm}\))\hspace{-1mm}\Psi(m,t). 
        \eea 
         We can now calculate the phase shift 
         of the wave function $ \Phi_{ADB}(m,t) $ at the end of the  adiabatic cycle: 
           \bea 
           \arg\((\frac{\Phi_{ADB}(m,T) }{\Phi_{ADB}(m,0)} \)) &=&
         \arg\(( \frac{ \Psi(m,T)}{\Psi(m,0)}\)) +\nonumber\\
       && \hspace{-22mm} \frac{1}{\hbar}\((\gamma(m, T)  - \int _0^T d t \, \En(m,B(t),E(t)) \))
           ,\nonumber \\
       && \hspace{-22mm}=  -\frac{1}{\hbar}\int _0^T d t \, \En(m,B(t),E(t)) + \beta(m),
        \label{ADBPhi}
             \eea 
        where we have  made  Berry's phase $\beta(m)$ stand out on the r.h.s   of the above equation.  
  Using the equation   (\ref {Shifgau}),
  one gets    immediately the basic formula
   giving     $\beta(m)$  in terms of the instantaneous wave functions $\Psi(m,t) $:
   \begin{eqnarray}
\beta(m)&= & \int_0^T   dt \bra\Psi(m,t)\vert  \, i\, \frac {\partial}{\partial t} 
\Psi(m,t)\ket+ \phi(m),   \nonumber \\  
 \phi(m) &=&\arg\((\Psi(m,T)/\Psi(m,0) \)). 
\label{formphas}
\end{eqnarray}

It is crucial to note that the  ``dynamical phase
" $\phi_D(m) $ 
and $\beta(m)$ obey different scaling laws under the 
transformation, $ B \rightarrow   \xi  B $,  involving an arbitrary real parameter $\xi$, while keeping invariant the Euler 
angles and the dimensionless parameter $\lambda$. If one remembers that 
$\En(m,B(t),E(t)) =\gamma_S  B \,  { \cal E } (m, \lambda(t) ) $, one sees immediately 
that  $  \phi_D(m)$  is multiplied  by $  \xi $,  while $\beta(m)$, 
being geometric, is unchanged. 
In principle, this  scaling  difference could be   used 
to separate the two phases. However, a more practical way to isolate
 Berry's phase consists in measuring the  phase for a given  adiabatic cyclic evolution  
and that associated 
with  the  ``image"  circuit  obtained by performing on 
     the Hamiltonian parameters the  transformations:
       $ \alpha(t) \rightarrow - \alpha(t) \, ,\,\varphi(t) \rightarrow - 
\varphi(t) $, while keeping the other two unchanged.
 The two competing  phases  are transformed as 
        $  \phi_D(m) \rightarrow \phi_D(m)   \, ,\,  \beta(m) \rightarrow  - \beta
(m)$, so that the dynamical phase can be eliminated by subtraction.

 To end this mini-review, we would like to give, within the present
context,   a  simplified   description of  the topological  interpretation of 
the Berry's Phase, due to Simon  \cite{simon}. We have just shown that $\Phi_{ADB}
(m,t) $ is a physical state obeying the Schr\"odinger equation within the adiabatic 
approximation.  For our purpose, it is convenient to introduce the $ mathematical $   vector 
state  
 $ \Phi _{\parallel}(m,t)= \exp \((-i   \phi_D(m,t)) \)) \Phi_{ADB}(m,t) $ 
 and to calculate  the  differential form 
 $ \bra  \Phi _{\parallel}\vert d  \Phi _{\parallel}\ket =
 \bra  \Phi _{\parallel}(m,t)\vert  \frac{\partial}{ \partial t} \Phi _{\parallel}(m,t)\ket
\,  dt  $  taken along the adiabatic loop:
 \bea 
 \bra  \Phi _{\parallel}\vert d  \Phi _{\parallel}\ket &=&
  \bra  \Phi _{ADB} \vert -i \frac{\partial}{ \partial t}
    \phi_D(m,t) \Phi _{ADB} \ket \, dt  +\nonumber \\
  && \bra  \Phi _{ADB} \vert  \frac{1}{ i \hbar } H(t) \Phi _{ADB} \ket \, dt= 0 .
  \nonumber
  \eea 
   The evolution of the state  vector $  \Phi _{\parallel}$   along   the  closed loop  is  then  
said to satisfy  the ``parallel transport" condition $\bra  \Phi _{\parallel}\vert d  \Phi _
{\parallel}\ket =0$.  If  the state  
$  \Phi _{\parallel}(t)$    is injected  into the general formula for the Berry's phase, one  
immediately finds that 
\be
  \beta (m)=  \arg(\frac{\Phi_{\parallel}(m,T) }{\Phi_{\parallel}(m,0)} ),
  \label{phasepr}
  \ee
  in the case of parallel transport.
  
 Let us now give a rather elementary  introduction to  the mathematical concepts  
behind the above notion of ``parallel transport" applied to the  evolution of quantum 
states.  This arises rather  naturally from a  linear  fiber bundle  interpretation  of 
Quantum Mechanics. The  linear fiber bundle associated with the  quantum state space   is 
constructed  from the  ``base space"  $ E(\rho) $    of the ``pure " state density matrix
 $\rho = \vert\Psi \ket \bra \Psi\vert $.  Assuming for simplicity that the states $\Psi$ have  
unit norms, one finds that $\rho$  satisfies the simple  nonlinear relation $\rho^2 =\rho$. This  implies 
clearly that $ E(\rho) $ has  a non-trivial topology. The ``fiber" is  the one-dimensional  
space associated   with a definite  $\rho$.  The    vector states of  the fiber  are  given   by   
$ \Phi= \exp( i \phi ) \Psi $,  where $ \phi $ is an arbitrary phase and $\Psi $ a 
representative state  of the fiber.  The infinitesimal variation  $ d \Phi $ during the 
quantum cycle  is said to be  ``vertical"  if it takes place  along the fiber: 
 $ d \Phi_{V}  =   d \tau \, \Phi_{V}  $, where  $d \tau $  is an infinitesimal  $c$-number. 
Conversely,  it is called  ``horizontal" or ``parallel" if      
 $ \bra  \Phi_{\parallel} \vert  d  \Phi_{\parallel} \ket  =0 $. The fact that the  Berry's phase 
can be viewed as a displacement respective to  the  fiber, associated with $ \rho(0)= \rho(T)$ - in our 
case a phase shift  resulting from a parallel transport   along a closed loop  drawn  upon  
the base space $ E(\rho) $ - emphasizes its topological character. 
 
\subsection{The Berry's phase  as a loop  integral  in the Hamiltonian
   $  H(\vB(t),\vE(t))$  parameter space}

Our starting point is  the   formula (\ref{formphas})   giving Berry's phase for  quantum adiabatic 
cycles associated with the instaneous wave function   $ \Psi(m,t )$.
 
The fundamental property of  $\beta(m)$ is its  invariance under the  gauge 
transformation
 $\Psi(m,t) \rightarrow \exp  \((  \frac{i}{\hbar} f(t)  \)) \,\Psi(m,t)  $, 
 where $f(t)$ is an arbitrary real function. 
 The density matrix of an isolated quantum system has clearly the  
same  gauge invariance  property.
 The adiabatic approximation allows us to make a mapping of the Hamitonian parameter 
  space    onto the density matrix space. As a  consequence,    
 $\beta(m)$    could  also  be viewed  as a  line integral 
along a closed path drawn in the density matrix space.
\paragraph{A group theoretical derivation}

   To evaluate the expression (\ref{formphas}), it is convenient to  write $\beta(m)$ in terms of $\hat \psi(m,
\lambda)$:
\bea
&&\beta(m) -\phi(m)= \int_0^T dt \bra \hat \psi (m, \lambda(t)\vert i\,\frac 
{\partial}{\partial t}\hat \psi (m, \lambda(t) )\ket +  \nonumber \\
& &\hspace{-8mm} \int_0^T \hspace{-2mm}dt \,\bra \hat \psi (m, \lambda(t))\vert 
U^{-1}(R(t))  i\,\frac {\partial}{\partial t}\,\left ( U(R(t)) \right ) \vert
\hat \psi (m, \lambda(t) \ket. 
\label{Gtr}
\eea 
Since $ \vert \hat \psi (m, \lambda(t) \ket $ is a real vector in the $\vert S, m
\ket$ basis, the first term of the integral can be written as:
$$
\frac{i}{2}\frac{\partial}{\partial t}  \{\Sigma_{_{\mu=-S}}^{^ {\mu=S}}\;
[\bra S,\mu\vert \hat \psi (m,\lambda(t)\ket]^2\}\;.
$$
If the cyclic  condition  $ \lambda(T) = 
\lambda (0)$ is satisfied, the  line  integral of this term is zero.  
If now one uses the explicit form of $U(R(t))$ in terms of the spin operator $\vS$ 
(Eq. (\ref{UR}), it is easily  seen that  the second term of Eq.(\ref{Gtr})
 is a linear combination of time derivatives of the Euler  angles:
\be
 i \hbar U^{-1}(R(t)) \frac {\partial}{\partial t}\,( U(R(t)) =\dot{\alpha} \; D_
{\alpha} + \dot{\theta} \;
D_{\theta} + \dot{\varphi} \;  D_{\varphi} \,. 
\label{UdU}
\ee
Let us consider $D_{\alpha}$: 
\bea
&&D_{\alpha} = \exp{(\frac{i}{\hbar} S_z \alpha(t))}\cdot \exp{(\frac{i}
{\hbar} S_y \theta(t))}\cdot
 \exp{(\frac{i}{\hbar} S_z \varphi(t))} \times\nonumber\\
 &&\hspace{-3mm} \exp{(-\frac{i}{\hbar} S_z \varphi(t))} \cdot  
\exp{(-\frac{i}{\hbar} S_y \theta(t))}\cdot S_z \cdot\exp{(-\frac{i}
{\hbar} S_z
\alpha(t))} \nonumber. 
\eea
Since $S_z$ commutes with the exponential $\exp{(-\frac{i}{\hbar} S_z
\alpha(t))}$, $D_{\alpha}$ reduces to:
\be
  D_{\alpha}=   U^{-1}(R(t)) U(R(t)) S_z =  S_z . 
 \label{Dalpha}
\ee
 For the two  remaining terms such as $D_{\varphi}$ the calculation is not so simple. 
Using the fact that $S_z $ commutes  with
$\exp{(-\frac{i}{\hbar} S_z \varphi(t))}$, one can write 
\bea
 D_{\varphi}&=&\exp{(\frac{i}{\hbar} S_z \alpha(t))}\cdot \exp{(\frac{i}{\hbar} 
S_y \theta(t))}\times \nonumber \\
&&  S_z \cdot \exp{(-\frac{i}{\hbar} S_y \theta(t))}\cdot\exp{(-\frac{i}{\hbar} S_z 
\alpha(t))}. \nonumber
\eea
Using equation(\ref{GR2}), one can derive the commutation relation:
$$S_z\cdot \exp{(-\frac{i}{\hbar} S_y \theta(t))}=\exp{(-\frac{i}{\hbar} S_y 
\theta(t))} \cdot ( S_z
\cos{\theta} -  S_x \sin{\theta})\,.$$
After repeating a  similar operation to push $ \exp{(-\frac{i}{\hbar} S_z
\alpha(t))}$  to the left of $( S_z
\cos{\theta} - S_x \sin{\theta})$, one arrives  finally at the following 
expression: 
\be
D_{\varphi}= S_z \cos{\theta} + \sin{\theta} ( - S_x \cos{\alpha} + S_y \sin
{\alpha}) \,.
\label{Dphi}
\ee  
In a similar way one obtains  
 \be
  D_{\theta}=   S_y \cos{\alpha} + S_x \sin{\alpha}  ,
  \label{Dtheta} 
  \ee 
  but this term will  not lead to any contribution to $\beta $ because
   of the boundary condition: $\theta(T)= \theta(0)$.
Since the quantum average of $\vS$ relative to the state $\hat \psi(m,\lambda) 
$ has a single component along $\hat z$, we end up with  the  compact  expression:
\be
\beta(m) -\phi(m)=  \int_0^T p(m,\lambda(t))(\cos{\theta} \; \dot{\varphi} + 
\dot{\alpha})  dt \,.
\label{betaminusphi}
\ee

  We now have to calculate the phase shift
$ \phi(m)=\arg\left \{\Psi(m,T)/\Psi(m,0)\right\}$ appearing in Eq (\ref{formphas}). 
 A preliminary step is to rewrite  
 the unitary transformation   $ U(R(t))$  under a modified form:
 \bea
 \hspace{-3mm}U(R(t)) &= & V(t)  \, \exp\((-  \frac {i} {\hbar}\,S_z ( \varphi(t)  +\alpha(t))  \)), \nonumber \\ 
 V(t)  &= & \exp \((-  \frac{i}{\hbar}\ S_z  \, \varphi(t) \))  \times \nonumber \\ 
 & &  \,\exp \(( - \frac {i }{\hbar}\,S_y \, \theta(t) \))
  \, \exp\((\frac { i}{\hbar}\,S_z  \, \varphi(t) \)).
  \eea
The transformation law given by equation (\ref{GR2}) can be extended to any 
tensor operator $ S_i\, S_j \,S_k  \,... $ and consequently
 to any analytical function of $ S_y $. Using   the extended law,
  one can write $V(t) $ 
in the compact form:
\be
 V(t) = \exp \((  - \frac{i}{\hbar} \theta(t)  (  \cos \varphi(t)   \, S_y +
 \sin \varphi(t)  \,S_x) \)) .
 \ee 
Using the boundary conditions $\theta(T)= \theta(0), \lambda(T)=\lambda(0),  
 \alpha(T) = \alpha(0) + n_{\alpha} \pi$  and $  \varphi(T) = 
\varphi(0)+2 n_{\varphi} \pi,   $ one obtains  the relation: 
$ V(T) =V(0) $.  Introducing for  convenience 
the notation $  U( \hat z , u)= \exp\(( -\frac {i}{\hbar}\,S_z \,u \))$  it is then possible
 to rearrange  $\Psi(m,T)$ as follows:
Ê\bea
\Psi(m,T)& = & V(0)\,  U( \hat z , \varphi(T)  +\alpha(T )) \hat \psi(m, \lambda(0)) , \nonumber \\
&&\hspace{-18mm}=V(0)  \, U( \hat z , \varphi(0) +   \alpha(0))\, U( \hat z , \Delta \varphi +   \Delta\alpha )  \hat \psi(m, \lambda(0)) ,\nonumber \\
&&\hspace{-18mm}\text{with}\;\;\;\Delta \varphi+ \Delta\alpha=  \varphi(T)  +\alpha(T )-\varphi(0)-\alpha(0), \nonumber \\
& & \hspace{7mm} =  (2  n_{\varphi} +n_{\alpha})\pi .  
   \label{phim}
\eea
The next step is to prove that $ \hat \psi(m, \lambda(0))$ is an eigenstate  of
$ U( \hat z , \Delta \varphi + \Delta \alpha )$ using its  expansion
in terms of the  angular momentum state vectors  $\vert S, q\ket $, eigenstates of $S_z$ of eigenvalue $\hbar q$:  
\be
 \hat \psi(m, \lambda(0))=\sum _{\vert m- 2n \vert \leq S } C_{m,n}(\lambda(0)) \vert S, m-2 n \ket. \label{psihatexpand}
 \ee
Using equations (\ref{phim}), one sees immediately that, as a consequence of the 
quantum cycle boundary conditions,  each state $ \vert S, m-2 n \ket$ appearing in the 
above sum is an eigenstate  of $ U( \hat z , \Delta \varphi + \Delta\alpha )$ with the eigenvalue 
$\exp \((-i \,m \,\pi  ( 2 \,  n_{\varphi} + n_{\alpha} )  \))$.
 This  leads to the basic   relation connecting $\Psi(m,T)$ and $\Psi( m,0)$:
           \bea
 \Psi(m,T)&=&
   \exp \((  -i \,  m\,\pi  (  2 \,   n_{\varphi} + n_{\alpha} )  \))  \times \nonumber \\
  & & V(0) \,  U( \hat z , \varphi(0)  +  \alpha(0) ) \hat \psi(m, \lambda(0))   \nonumber \\
 & = &   \exp\(( -i\,m \,( \Delta \varphi+ \Delta\alpha     )  \))  \,\Psi( m,0).
     \eea
 One immediately obtains the phase shift $ \phi(m)$ appearing on the r.h.s 
of equation (\ref{formphas}):
\bea
\phi(m)  &=&\arg\left \{\Psi(m,T)/\Psi(m,0)\right\}\nonumber\\
&=& -m ( \varphi(T) -\varphi(0) + \alpha(T)-\alpha(0)) \nonumber \\
&=& -m \int_0^T ( \dot{\varphi}(t) + \dot{\alpha}(t)) \; dt \;. 
\label{phi}
\eea 
By combining the equations (\ref{betaminusphi}) and (\ref{phi}), we arrive
 at the expression of Berry's phase for an arbitrary quantum adiabatic 
cycle generated by the Hamiltonian $H (\vB(t), \vE(t))$ of equation (1) with
$\vE(t) \cdot \vB(t) =0$ 
\bea
\beta(m)&=&  - \int_0^T  [m- p(m,\lambda(t)) \cos{\theta(t)} ]\,\dot{\varphi}(t) \;dt
\nonumber \\
 &&    - \int_0^T  [m-p(m,\lambda(t))]\;  \dot{\alpha}(t) \;dt \,.    
  \label{betafinal}
\eea
\paragraph{The Berry's phase as a loop integral of an Abelian Gauge Field}

 The Berry's phase  $ \beta(m)$ given by equation (\ref{betafinal})  
 can be written as a loop integral around a closed 
 curve $ {\cal C }$ drawn in the parameter space    $ \vC P^2 $,
 using $ \lambda, \theta, \varphi, \alpha $ as coordinates:
\be 
\beta(m)=    -   \oint_{{\cal C }} \(( Ê ( m- p( m, \lambda)  \cos \theta ) dÊ\varphi + 
  (m-p(m,\lambda)) d \alpha  \)) .
     \label{betageom}
\ee
The above formula suggests that $ \beta(m)$  could be written as a loop line integral of  an Abelian gauge field $ A_{i}( x_j )$,  defined in  the parameter space.    
The non-vanishing  components of the gauge field candidate,   $ A_{\varphi}$ and $ A_{\alpha}$,  are   then   given by:
\be  
A_{\varphi}= - m+p( m, \lambda)  \cos \theta, \; \; \; \; \; \; 
A_{\alpha } = - m+p(m, \lambda).
\ee 
  Thus the  Berry's phase  looks like the Bohm-Aharonov phase \cite{Bohm}, for the Abelian gauge field $ (A_{\varphi},A_{\alpha)} ) $: 
 \be  
 \beta(m)= \oint_{{\cal C }}( A_{\varphi}\, d\varphi +A_{\alpha }  d\alpha  ). 
 \ee
 What remains to be proved  is that the above expression is indeed invariant under the  gauge transformation  of the Abelian field:
 \be
 A_{\varphi}\rightarrow A_{\varphi} +\frac{\partial}{\partial \varphi } g(\varphi, \theta, \alpha,   \lambda),   \;  \; \;         A_{\alpha}\rightarrow A_{\alpha} +\frac{\partial}{\partial \alpha }g(\varphi, \theta, \alpha,   \lambda) .      
 \ee 
 The crucial point is that $ g(\varphi, \theta, \alpha,   \lambda) $ {\it must  be a single-valued  function of   $\vB $ and $ \vE\otimes\vE$, like the gauge field itself}. This implies  that 
$ g(\varphi, \theta, \alpha,   \lambda) $  must  be a periodic function of $ \varphi$ and
 $\alpha$,  with periods respectively of  $ 2 \pi  $ and $\pi$.  Remembering the 
 boundary conditions (\ref{boundcond}) for the quantum cycle  ${\cal C }$ , one finds immediately  that 
 the gauge contributions to $\beta(m)$,   
 $  g(\varphi(0) + 2 n_{\varphi}\, \pi , \theta(0), \alpha(0)+n_{\alpha}\, \pi  ,   \lambda(0))-
 g(\varphi(0) ,  \theta(0), \alpha(0) , \lambda(0))$, do indeed vanish.
 This completes the identification of $\beta(m)$ as a Bohm-Aharonov phase.
  \begin{figure}
  \centering\includegraphics[  ]{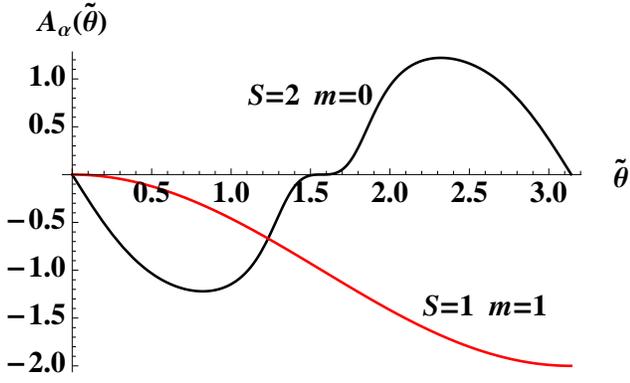}
\caption{Representation of the gauge field component $A_{\alpha}(\tilde{\theta})$ versus the associated 
$\tilde{\theta}$ parameter of the non-linear 
Hamiltonian in the two cases $S=1$ (red curve) and $S=2$ (black curve).
The difference between the two graphs illustrates the fact that for $S>1$  there is no longer 
 a one to one correspondence between the Hamiltonian parameter space and the density matrix space
 upon which is drawn the quantum cycle.  The non-linear spin Hamiltonian generates a new geometry revealed only by spins larger than one.}
 \label{geom}
 \end{figure}
\paragraph{ Berry's phase geometry  generated by non-linear spin
Hamiltonians for $ S > 1$}

For the sake of simplicity, in the next sections we shall concentrate on Berry's cycles where $\lambda $ and $\alpha$ are the only
time-varying parameters. The question then arises as to whether we shall not
lose in this way most of the new features introduced with the
non-linearity of $H(t)$. In order to make clear the cause for our concern, let us  perform the change of variables upon the two left-over parameters
which map the associated  2D manifold onto a 2D sphere:
$\lambda=-2 \cot (\tilde{\theta })$ with $  0<  \tilde{\theta } < \pi $ and
$  \tilde{\varphi}= 2\, \alpha $.
In ref. \cite{bou1} we have given an explicit form of the $ \vC P^2 $
metric obtained with this kind of coordinates. For the sub-manifold associated  with $
\tilde{\dot{\varphi}}=\tilde{\dot{\theta}}=0$
it turns out that this  metric, 
 $ ds^2 =\frac{1}{4} \left(d \tilde{\varphi }^2 \sin ^2\,\theta
   +d \tilde{\theta }^2\right)$
 is identical to the one associated  with a 2D sphere having a radius 1/2
(this factor results from the choice:
$  \tilde{\varphi}= 2\, \alpha $.)

To answer the question, we can now consider the component of
the gauge field $ A_{\alpha}$ for two different cases. For $S=1, m= 1$, the
gauge field is given by the same expression as the gauge field associated
with a linear Hamiltonian
$ A_{\alpha}(\tilde{\theta}) = \cos \tilde{\theta} -1$. However, the
situation is totally  different for the case $S=2, m=0$, where $
A_{\alpha}(\tilde{\theta})$ exhibits the peculiar shape shown in Fig.2.
This implies that the geometry involved in Berry's phase for the 
sub-manifold  differs from its usual interpretation in terms of the solid
angle defined by a closed loop drawn upon a 2D sphere. The origin of this 
phenomenon lies in the fact that the density matrix space for $S=2$ is
not $ \vC P^2 $  but rather $ \vC P^4 $. Unlike  the case $S=1$, there is  no
longer a one to one correspondence between the parameter space and the
density matrix space. It is thus not surprising  that the geometry
involved in Berry's phase for a non-linear Hamiltonian should  differ
from the linear case, even for the simple cycles involving  only the variations of 
the two parameters $ \lambda $ and $ \alpha $.
   \section{Non-adiabatic corrections in Berry's cycles using the rotating frame Hamiltonian }   
  For quantum cycles of finite duration $T$,  
  the time derivatives of the Hamiltonian  parameters cannot take arbitrary small values.
  For instance, in the experiments discussed in a separate paper \cite{bou3}, the quadratic spin coupling   generated by an ac Stark effect induces an atomic instability when  $\lambda \approx 1$.
 As a result, the measurement of Berry's phases  studied  by interferometry
   experiments on  alkali atoms will require a good control upon   
   the non-adiabatic corrections arising from the time dependent external $ \vB,\vE $ fields.
  
 We shall consider separately the effect coming from the periodic parameters $\alpha $ and $\varphi$ in subsections V.A and V.B,  and the non-periodic one $\lambda$ in subsection V.C. 
  Our method involves the time-dependent Schr\"odinger equation in the rotating frame.   We shall use an adiabatic approximation within this frame, the slowly varying parameters being, then, the time derivatives of the periodic parameters $\dot \alpha$ and $\dot \varphi$. 
 
 We set  $\Phi(t) = U(R(t)) \wt \Phi(t)$, and we write the   
wave equation  relative  to  $\wt \Phi(t)$ in the rotating frame:
\bea
&&\((i \hbar \frac{d}{dt}- \hat{H}(B(t),E(t)) \))  \wt \Phi(t) =\nonumber\\
 &&\hspace{-5mm}- U^{-1}(R(t))\,i \hbar \frac{d}{dt }\, U(R(t))\,\wt \Phi(t) =\nonumber \\
& &\hspace{-10mm} -( D_{\alpha }\,\dot\alpha + D_{\varphi}\,\dot\varphi +D_{\theta}\,
\dot
\theta)\,\wt \Phi(t)  =
  \gamma_S \, \vS  \cdot\Delta \vB(t) \,\wt\Phi(t).
 \eea
 The rotating frame Hamiltonian
 can be written as
 \be
 {\wt H}(t)= \hat{H}(B(t),E(t)) + \gamma_S \vS \cdot \Delta \vB (t),
 \ee
 where   $\Delta \vB(t) $ is   an additional effective magnetic field 
 generated by   the Coriolis effect. It can be decomposed into 
 a longitudinal component $ {   \Delta \vB}_{//}(t)  $ and a transverse one $ {\Delta \vB}_{\perp}(t)$.
  Using  the explicit expressions 
of $D_{\alpha }$ and  $D_{\varphi}$ in subsection III.B and  
assuming  here -for the sake of simplicity-
that $\dot\theta=0 $,  one arrives at  the following expressions for 
 of the effective field:  
 \bea 
   {\Delta \vB }_{//}  & = &   - \gamma_S ^{-1} \, ( \cos\theta\; \dot 
\varphi+ \dot \alpha ) \,\hat z   \nonumber \\ 
  {\Delta \vB}_{\perp}& =& - \gamma_S ^{-1} \,   \sin \theta \; \dot  
\varphi \,  ( - \cos\alpha \,\hat x +\sin\alpha \,\hat y )   \nonumber \\
 &=& \gamma_S ^{-1} \,   \sin\theta \; \dot  \varphi \, {\cal  R} ( \hat 
z , - \alpha )  \, \hat x .
        \eea  
 The role of those two components are going to be examined separately.       
\begin{figure*}
\includegraphics[height=8.0 cm]{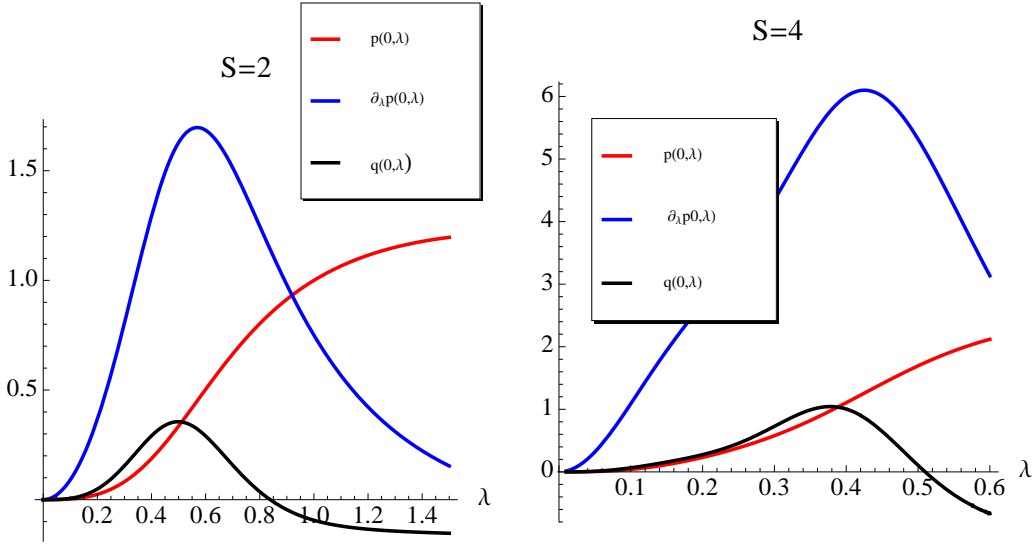}
\caption{ \small  Comparison of   Berry's phase 
   to its lowest-order non-adiabatic corrections, when $ \alpha  $ is the 
only time-dependent Euler angle, for a quantum cyle involving  the spin configurations  
$S=2$, or 4 and $m=0 $.
   The Berry's phase is given by $ \beta(0)= \int _0 ^T d t \,  
p(0,\lambda(t) ) \dot \alpha (t)  $. 
Using the results of   subsection V.A, one finds 
that the non-adiabatic    correction is given by  :
     $\Delta_{//}^{(2)}\beta(0) =  
     \int _0 ^T d t \,  q(0,\lambda(t) ) \dot \alpha (t) \, 
     (\frac{\dot \alpha (t)} {\gamma_S \, B(t) })^2 $. 
     The explicit expression of $q(m,\lambda ) $  in terms of the second- and 
third-order  derivatives of 
     ${\cal  E}(m,\lambda)$   with respect to $\lambda$ can be  found  in 
Eq. (\ref{ql}). One sees clearly that, for $S=2$, if the parameter $\lambda $ is required 
to stay within the small interval  $ 0.8\pm 0.2 $, then  
 $\Delta_{//}^{(2)}\beta(0) /\beta(0) $ will  be $ < 10^{-3} $   provided  
    $   \dot \alpha (t)  $   is  kept    smaller than one tenth  
 of the Larmor  angular frequency $\gamma_S \, B(t)$. Similarly, for $S=4$, there is a value
of $\lambda$, for which $q(0,\lambda)$ cancels.  The partial derivative
$ \frac{\partial}{ \partial \lambda } p(0,\lambda) $
 is  governing the response of $\beta(0,\lambda)$, say, 
 to a small variation of the $ B$ 
field, are nearly at their   maximum when $ q(0, \lambda) $ is close to 0.  
Note also the  fourfold   increase of the sensitivity for $ S=4 $ compared to $ S=2 $.}
\label{fig3}
\end{figure*}
 \subsection{The Berry's phase and its non-adiabatic corrections  
  involving the longitudinal effective magnetic field $( B(t) + \Delta B_{//}(t)) \,\hat z$}
  We consider, first, the part of the Hamiltonian governing the 
evolution of $ \wt \Phi(t)$  associated 
 with $ {\Delta \vB }_{//} $,  which is  the sole present if $\dot\varphi = 0$.  This part can be written very simply in terms of  $ \hat H(B,E) $
 \be
 \wt H_{//}(t) =  \hat H(B(t) + \Delta B_{//}(t), E(t) ). \; \; \label{Htdparal}\\
   \ee 
To proceed  it is convenient  to introduce the dimensionless parameter  $\eta$ : 
\be  
  \eta= - \frac {\Delta B_{//}}{ B}=  
    \frac{\cos{\theta} \,\dot \varphi +\dot \alpha} {\gamma_S \, B }. 
  \label{formeta}
\ee 
   Using $\wt B= (1-\eta) B$, the eigenvalues of  $ \wt H_{//}(t) $ are simply  
\bea 
 \hspace{-3mm}\En(m, B +\Delta B_{//} , E )& =&\gamma_S \wt{B} \hbar\, \mathcal{ E}( m, \tilde{\lambda }) 
 \nonumber \\
  &=&  \gamma_S \, B \hbar\,
(1- \eta )  \,\mathcal{ E} (m , \frac{\lambda} {1- \eta} )  
  \eea
Since the quantum cycles generated by $\wt H_{//}(t)$ are devoid of any topology, we can expect Berry's phase relative to the laboratory frame to be buried inside the dynamical phase of $\wt H(t)$: 
\be
\tilde  \phi_{D \, // }(\lambda,\eta,m)=  - \int _0 ^T dt \,  \gamma_S \, B
(1- \eta ) \mathcal{ E} (m , \frac{\lambda} {1- \eta} ).  \label{phiDpar}
\ee
 Indeed, it is easily seen that Berry's phase is given by the first-order  contribution
 in the   $\eta$-series expansion of the above integral :
 \bea
 \tilde  \phi_{D \, // }^{(1)}(\lambda, \eta, m)&= & \int _0 ^T dt \,  \eta \, \gamma_S \, B
   \((\mathcal{E}(m,\lambda)   - \lambda \frac {\partial { \mathcal E}(m,
\lambda)}{\partial \,\lambda}\)), \nonumber \\
& =&  \int _0 ^T dt \; p(m, \lambda) \,(\cos{\theta} \,\dot \varphi +\dot \alpha ), \nonumber  \\
& = &  \,\beta(m) -\phi(m), \label{betapar}
\eea
 where we have used equation (\ref{polarvsla})  of sec. III. The phase $\phi(m)$ is recovered by returning to the laboratory frame. The even-order contributions $\propto \eta^{2p}$ are eliminated by subtracting term by term those coming from image circuits associated to opposite signs of the $\eta $-parameter ({\it i. e.}  $\varphi \rightarrow -\varphi$ and $\alpha \rightarrow -\alpha$).  

We now show that it is possible to find a magic value of $\lambda = \lambda^{\star}(\eta)$ that leads to a cancellation of all the contributions of the dynamical phase $\propto Ê\eta^{2p+1}$. More precisely let us define 
\bea
\Delta \beta_{//}\hspace{-2mm}& =&\hspace{-2mm}
 \frac{1}{2} \(( \tilde  \phi_{D \, // }(\lambda,\eta,m)-\tilde  \phi_{D \, // }(\lambda,-\eta,m) \))\hspace{-1mm} - \hspace{-1mm}
 \beta(m) \hspace{-1mm}+\hspace{-1mm}\phi(m) \nonumber \\
 &=&\int_0^T dt(\cos{\theta} \,\dot \varphi +\dot \alpha) \, \Delta p(m,\lambda,\eta),\label{Deltabetapar} \\
 &&\hspace{-12mm}\text{where}\;\;\; \Delta p(m,\lambda,\eta)= \nonumber \\
 &&\hspace{-7mm}\frac{1}{2 \eta }  \((\eta +1) \mathcal{E}(m,\frac{\lambda }{1+\eta} ) - 
 (\eta \rightarrow - \eta)   \))-  p(m,\lambda).   \label{deltap}
\eea  

The above formula seems to suggest that  the integrand in the expression 
giving
  $\Delta \beta_ { //}(m) $  is singular when $ \eta =\pm  1$.  In fact it is easily
  seen that this is not the case.  Indeed, the ratios  
   $ \mathcal{E}( m,\lambda)/\lambda$
  converge  towards the  eigenvalues  of $ (S_{x}/\hbar)^2 \,: \,0,1,4 ...  
S^2$ , as $ \lambda \rightarrow  \infty $.  
  As a preliminary step, let us   consider the lowest-order expansion  of the r.h.s of 
  $\Delta p(m, \lambda, \eta)$  with respect to $ \eta $ 
 \bea 
   \Delta \beta_ { //}^{(2)} (m) &= &\int _0 ^T dt \, 
      \left (\cos{\theta} \,\dot{\varphi} + \,\dot \alpha  \right )
      q(m,\lambda) \, \eta^2 ,
  \nonumber     \\
     q(m,\lambda) &=&\frac{-1}{6}\((  \lambda ^3\,{\cal E }^{'''} (m,
\lambda) +    3 \,\lambda^2\, {\cal E }^{''}(m,\lambda) \)) .
\label{ql}
\eea  
 Fig.\ref{fig3} shows the variations of   $  q(m,\lambda) $ versus $ \lambda $ in the cases $ S= 2$ and 4, for $ m=0$. 
  A  quite remarkable effect, clearly visible on  Fig.\ref{fig3},  is the vanishing of $ q(0, 
\lambda $)  for a particular value of  $\lambda$ , in the vicinity of  the maximun of
  $\frac{ \partial }{ \partial \lambda} p(0,\lambda) $.  This  should allow for  rather  
accurate determinations of   $\beta(0)$, as explained in the caption  of Fig.\ref{fig3}. 
 
In the case where $\alpha $ is the sole time-dependent parameter we have found, that 
a similar property holds in fact to all orders  
   in  $ \eta $. More precisely, we have shown that there exists for 
  a given value of   $ \eta $  a  ``magic" value $\lambda^{\star}(S,\eta ) $
such that the non-adiabatic correction to $p(m,\lambda)$ cancels exactly.  
\be 
\Delta p(0, \lambda^{\star}(S,\eta ), \eta)=0.
\label{lamagic}
\ee 
   The magic values can be accurately represented 
by the polynomial fits given below:
\bea
\lambda^{\star}(2,\eta )&=&0.838213 - 0.0837823 \eta^2 - 0.0431478 
\eta^4 - \nonumber\\
&&0.0231887 \eta^6 - 
 0.0207986 \eta^8  \nonumber\\
\lambda^{\star}(4,\eta )&=&0.509982 - 0.0900927 \eta^2 - 0.0349985 
\eta^4 - \nonumber \\
&&0.0436495 \eta^6 + 0.0373634 \eta^8
\label{lamagfit}
 \eea
 As it is apparent upon the above formulas,  the ``magic" values $\lambda^{\star}(S,\eta ) $
 are slowly varying functions of $ \eta $ within the interval 
 $ 0\leq \eta \leq 0.5 $.
 We have verified that  they both  satisfy  the  equation (\ref{lamagic}) with 
 a  precision better  than $ 3 \times  10^{-7}$. 
 In conlusion, the above results   open the road to  measurements 
 of the Berry's Phase  $\beta(0)$ for  $ S= 2 $ and $S=4$, for  
quantum  cycles where $ \alpha $ is the sole varying Euler angle, 
 under conditions  where the {\it non-adiabatic  corrections 
 can be kept below the one  ppm  level.}
  \subsection{ Non-adiabatic corrections induced by 
   the effective transverse magnetic field $ {\Delta \vB}_{\perp}(t)$ }
   This subsection   is divided into two parts.
   In the first  we  calculate the  corrections to the 
   Berry's phase of the laboratory frame that are odd 
    under $ \eta $-reversal. These arise from  the  modification of the dynamical 
    phase induced by $\Delta \vB_{\perp}(t)$.
   Then we turn to the small Berry's phase in the rotating frame  
   associated with  the non-trivial topology
   of the quantum cycles. It arises from the 
   time variation of the effective field $ B(t) \,{\hat z} +\Delta \vB(t)$ acting in the rotating frame. Let us  stress again an important feature of the  transverse
   contribution, $\gamma_S  \vS \cdot  {\Delta \vB}_{\perp}(t)   $  to ${\wt H}(t) $. Being 
   proportional to $ \cos \alpha S_x  - \sin \alpha S_y $,  it mixes states of opposite   $m$-parity which can  be nearly degenerate, unless severe restrictions are imposed 
   upon the domain of variation of $ \lambda $. These constaints are directly read off from Fig.1. For  $S=4, m=1, 2$ and 3, only negative values of $ \lambda$ are allowed, 
     $- 2 \leq \lambda  \leq 0 $. 
   For  $S=2, m=0$  a larger interval  can be used:
     $ \vert \lambda \vert \leq  1 $. 
   \subsubsection{Correction to the dynamical phase}
   It is convenient to introduce the dimensionless parameters $ \mu $ and $\wt \mu$ and the dimensionless operator $ \vSig = \hbar^{-1} \vS$:
   \be 
   \mu=- \frac{{\Delta B}_{\perp} }{B}=  \frac{ \sin \theta \, \dot \varphi }{\gamma_S \, B},
   \;\;\;\;\wt \mu= \frac{\mu}{1-\eta}.
   \ee
   We rewrite $ \wt {H}(t) $ as follows:
   \be
   \wt {H}(t)= \gamma_S  \wt{B}\hbar\, ( {\cal H}(\tilde \lambda ) -
          \tilde{\mu} \,   (- \cos \alpha \, \Sigma_x +\sin \alpha \, \Sigma_y ) ) .  
   \label{Htilde}
 \ee
  To proceed, it is convenient to introduce the perturbation expansion  of 
  the eigenenergies of  $ \wt {H}(t) $:
  \be 
  \wt{E}( \tilde{\lambda} , \tilde{\mu} )=  \gamma_S  \wt{B}\hbar\, (  \mathcal{E}(m, \tilde{\lambda })+
   \tilde{\mu}^2 \,  \mathcal{E}_{\perp}^{(2)} (m, \tilde{\lambda })+\mathcal{O}( \tilde{\mu}^4 ) ). 
  \ee 
 The second order energy shift  caused by the  $\wt \mu $-contribution to  $ \wt {H}(t) $ 
 is given, to all orders in $ \eta $, by:
    \be
{\cal E}^{(2) }_{\perp}(m,\tilde{ \lambda})  = 
           \hspace{-4mm}\sum_{\vert n-m \vert odd} \hspace{-4mm}              
   \frac{ \vert  \bra \hat \psi (n, \tilde \lambda) \vert \cos{\alpha}\, \Sigma_x 
- \sin {\alpha} \, \Sigma_y  \vert \hat \psi (m, \tilde \lambda)\ket \vert^2}
{  \mathcal{E}(m, \tilde{\lambda })-\mathcal{E}(n, \tilde{\lambda })} . 
\label{2ndOener}
        \ee
          The symmetry properties  of $H(\lambda)$ (Sec.II)  have two consequences i) only $\vert n-m\vert$-odd terms contribute to the sum and ii) the cross terms involving $\Sigma_x \Sigma_y$ cancel out. 
This means that there is no contribution proportional to $ \sin{\alpha} \cos{\alpha}$. We are left with the terms prop to ${ \cos{\alpha}}^2$ and ${\sin{\alpha}}^2$. We shall assume that the velocity $\dot \alpha$ is a  slowly varying function of $t$ during the cycle, so that ${ \cos{\alpha}}^2$ and ${\sin{\alpha}}^2$ can be replaced by their average values of 1/2. 
 
 At this point we have found  convenient to solve   numerically two auxiliary problems, namely the search  of  the eigenstates  of the two Hamiltonians: 
 $$
 {\cal H}_{i}(\lambda,\mu) = {\cal H}(\lambda) - \mu \Sigma_{i} \;\;\; \text{for} \;\;\; i=x \;\;\;\text{and} \;\;\; i=y. 
 $$  
Let ${\mathcal E}_{x,y}(\lambda, \mu)$ be the individual eigenenergies of ${\mathcal H}_{x,y}$ which are even functions of $\mu$. To lowest-order in $\mu$, they take the form: 
\be
 \mathcal{ E}_{x,y} ( \lambda,\mu ) =   \mathcal{E}(\lambda)  +    \mu^2 \, \mathcal{ E}^{(2)}_{x,y}(\lambda)  +\mathcal{O}( \mu^4 ),  
 \ee
where ${\cal E}^{(2)}_{x,y}(\lambda)$  are obtained by an interpolation towards $\mu=0$.
 The overall second order energy  shift of Eq.(\ref{2ndOener}) can be expressed as:
\be
\mathcal{ E}^{(2) }_{\perp}(m,\tilde{ \lambda}) =\frac{1}{2} (Ê{\cal E}^{(2)}_{x}(\tilde{\lambda}) + {\cal E}^{(2)}_{y}(  \tilde{\lambda} ) ).  
\ee
We have now all we need to write  the contribution to the  dynamical phase associated  with  $ \wt{H}(t) $ to 2nd-order in $ \mu $ and all orders  in $ \eta $:
\bea
\phi^{(2)}_{D \perp}& =&  - \int _0 ^T dt \,\gamma_S  \, \wt{B} \,
\tilde{\mu}^2 \,\mathcal{ E}^{(2) }_{\perp}(m,\tilde{ \lambda}) ,\nonumber \\
  &=&  \int _0 ^T dt  \,\gamma_S  \, B \frac  {\mu ^2} {1-\eta}	\mathcal{ E}^{(2) }_{\perp}(m,\frac{\lambda}{1-\eta} ).
\eea 
In practice, we shall concentrate on  the  contribution  to the  dynamical  phase  
relative to  $ \wt{H}(t) $   of the order of  $ \mu^2\,\eta $,
which is easily derived from the above equation. In order to highlight the close connection 
with Berry's phase we replace $ \eta $ in the integrand by its explicit expression 
 of Eq.(\ref{formeta}):
 \bea
 &&  \phi^{(2,1)}_{ D \,\perp }(m) =   \int _0 ^T dt   \,
     \mu^2\,  ( \cos{\theta} \,\dot \varphi +\dot \alpha )     {p}^{(2)} (m,\lambda), \nonumber \\
&&   \text{with}\;\;\;    {p}^{(2)} (m,\lambda) = (1+ \lambda\, \frac{ \partial}{\partial \lambda }) 
    \, \mathcal{E}^{(2) }_{\perp}(m,\lambda) ).
    \label{coradiaT1}
 \eea 
    Despite  the sign  difference  in the $\lambda $ derivative, our notation  underlines  the analogy 
 of the above correction    with   the  Berry's phase, but the presence 
 of the factor  $ \mu^2 \propto  {\dot \varphi }^2$    spoils  its geometric character. 
   The quantity  $ {p}^{(2)} (m,\lambda)$   
   has been plotted in Figure \ref{coradiab} for the 
   cases $S=2, m=0 $ and $m=-1$ over the interval $0.7<\lambda<1.2$. 
  As expected, the  non-adiabatic corrections are  $ \lesssim \mu^2\ $ for $m=0,-1$.  We have performed
 a  similar  computation for $ m=1$ and found, in  constract, 
that   $ {p}^{(2)} (m=1,\lambda)$ \emph{ blows up to values }  $ \gtrsim  100 $.
 \begin{figure*}
\includegraphics[height=7 cm]{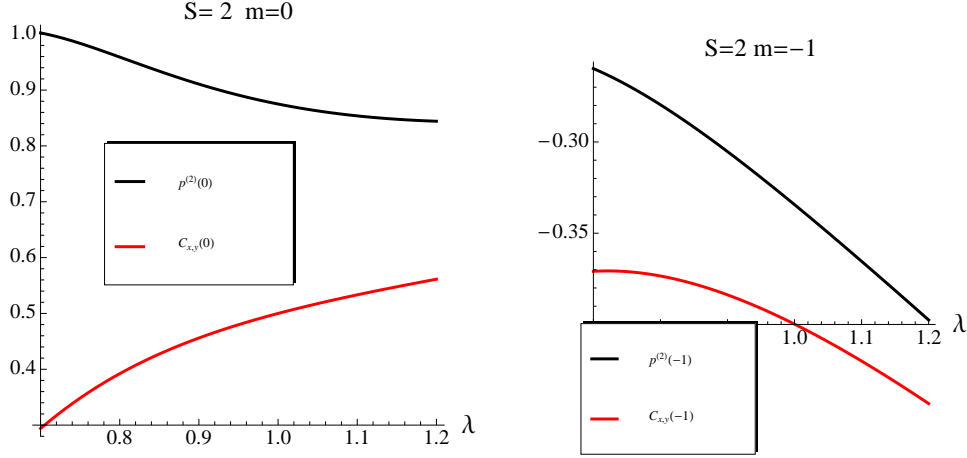}
\caption{\small Two non-adiabatic corrections to Berry's phase associated with 
the effective transverse magnetic field   $ {\Delta \vB}_{\perp}(t)$ acting in the rotating 
frame. The first,   
$\phi^{(2)}_{D \perp}=\int _0 ^T dt   \,
     \mu^2\,  ( \cos{\theta} \,\dot \varphi +\dot \alpha ) p^{(2)} (m,\lambda)$,  
is the  modification of the dynamical phase of $\wt H(t)$ associated with  $ {\Delta \vB}_{\perp}(t)$. 
The second,
$ \Delta^{(2)}  \beta_{\perp} (\tilde \lambda,\tilde\mu) =
 \int_0^T dt \, {\tilde \mu}^2 \dot  \alpha \,{C}_{x,y}(m,\lambda)$, is  
 Berry's phase in the rotating frame induced by the rotation of the transverse field,  
  ($\mu=\frac{ \sin \theta \, \dot \varphi }{\gamma_S \, B}, \,\wt \mu= \frac{\mu}{1-\eta}$).                    
Note that both corrections are not geometric phases because of 
 the presence of the factor $\mu^2 $. The results for the cases $S=2, m=0$ and -1 are of the same order of magnitude. They can be  used 
to  make Berry's phase determination precise at the order of $\mu^4$. }
\label{coradiab}
\end{figure*}
   \subsubsection{The Berry's phase associated with $\Delta \vB_{\perp}$}
   The full  Hamiltonian 
  $\wt H(t)=  \wt H_{//}(t)+ \gamma_S \vS \cdot {\Delta \vB}_{\perp}$, 
   generates quantum cycles endowed  with a non-trivial geometry
   since  the effective transverse magnetic field ${\Delta \vB}_{\perp}$
   rotates about the $ \hat{z}$ axis with the angular velocity $ \dot{\alpha} $.
   As above we  shall limit ourselves to the second order contribution with respect to $\mu$. 
   We proceed by writing the eigenfunctions of $\wt{ \cal H}(t)$ 
   in  first order expansion with respect to the parameter $\wt \mu $: 
    \be \hat{\psi}(m,\tilde{\lambda})-    \tilde{\mu} \, \hat{\psi}^{(1)}(m,\tilde{\lambda} )+ 
 \mathcal{O}( \tilde{\mu}^2 ).   
   \ee 
 In a way similar to what we did before,
  it is convenient to introduce the first order expansion of the 
 eigenfunctions 
$ \hat{\psi}_{x,y}(m, \lambda ) $ of $ {\cal H}_{x,y}(\lambda,\mu)  $.
$$ 
 \hat{\psi}^{(1)}_{x,y}(m,\lambda )=
  \sum_{\vert n-m \vert odd} 
\frac{   \bra \hat \psi (n, \lambda)\vert  \Sigma_{x,y} \vert \hat \psi (m, \lambda)\ket}{
  \mathcal{E}(m, \lambda )-\mathcal{E}(n, \lambda)}
 \hat\psi (n, \lambda). \nonumber
$$
 By using Eq.(\ref{Htilde}), one  finds that the     first-order contribution to    the  
 $\wt H(t) $  eigenfunction  (divided by  $ \tilde {\mu} $)  can be written as:
   \be
   \hat{\psi}^{(1)}(m,\tilde{\lambda})=
  - \cos \alpha \;\hat{\psi}^{(1)}_{x}(m,\tilde{\lambda}) +
   \sin \alpha \;\hat{\psi}^{(1)}_{y}(m,\tilde{\lambda}).
   \ee 
 It is then easily seen  that Berry's phase appears only to second order in  $ \mu $ and  
 involves the familiar time derivative product:  
 $ \int_0^T dt  \bra \hat {\psi} ^{(1)}(m, \tilde \lambda) 
      \vert i \, \hbar \frac{\partial}{\partial t}  {\hat \psi} ^{(1)}(m, \tilde \lambda) \ket$, 
which is proportional to $ \dot \alpha $. If the other parameters in the integrand vary slowly during 
the closed cycle, the only terms surviving  are those proportional to $ \cos^2 \alpha $ and 
$ \sin^2 \alpha$ which can be replaced by their average values of 1/2. This  
 leads to the following correction to the Berry's phase
\be
\Delta^{(2)}  \beta_{\perp} (\tilde \lambda,\tilde\mu) =
 \int_0^T dt \, {\tilde \mu}^2 \dot  \alpha \,
\Im \{\bra \hat{\psi}^{(1)}_{y}(m,\tilde{\lambda}) \vert 
\hat{\psi}^{(1)}_{x}(m,\tilde{\lambda}) \ket \}, 
\label{coradiaT2}
\ee
a result valid to all orders in $\eta$.  It is important to note that, unlike the 
case of the dynamical phase (Eq. 72), this correction invoves the product of $\tilde \mu^2 $ by 
$\dot \alpha $. Therefore,  it is the contribution odd in $\eta $ that is eliminated by 
the parameter reversal, so that the dominant contribution corresponds to the limit
 $ \eta  \rightarrow  0 $  { \it i.e }
  $ \tilde \lambda \rightarrow \lambda \, , \,  \tilde\mu  \rightarrow \mu   $. 
   The quantity  ${C}_{x,y}(m,\lambda)=
\Im \{\bra \hat{\psi}^{(1)}_{y}(m,\lambda) \vert \hat{\psi}^{(1)}_{x}(m,\lambda) \ket\} $  
 is plotted together with $p^{(2)}(m,\lambda)$ in Fig.5,  leading to the same  conclusions 
 as those reached at the end of the previous paragraph.
 
 \subsection{Non-adiabatic corrections associated with ramping-up the quadratic interaction}
 \begin{figure*}
\includegraphics[height=10 cm]{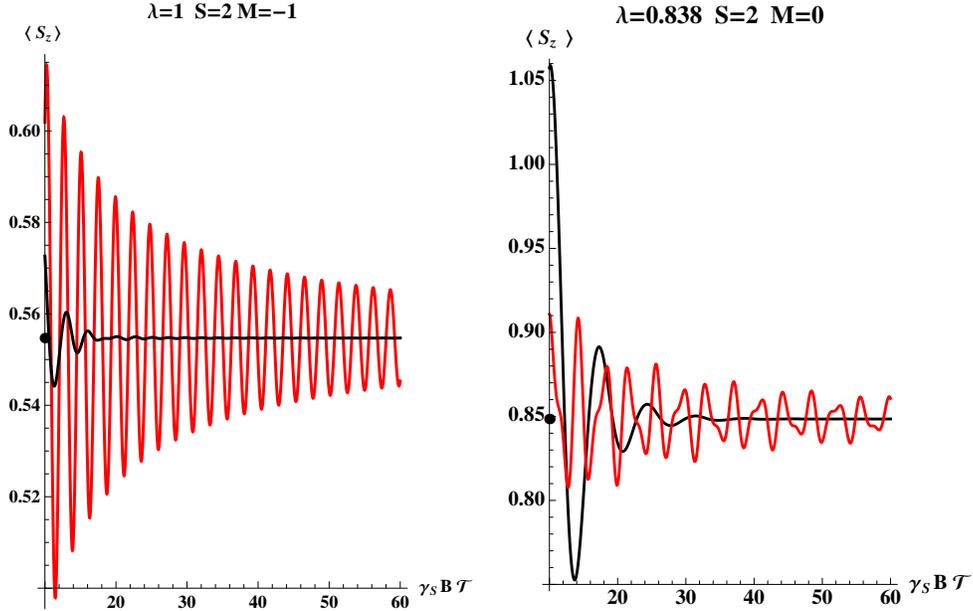}
\caption{\small 
Effect of taming the non-adiabatic oscillations generated by a too abrupt application of the quadratic spin interaction,  thanks to the use of a Blackman pulse-shaped time-derivative, oscillation taming (OT) procedure.  The diagram shows non-adiabatic corrections  to $ \langle S_z \rangle  $ at the end of the transition from $ \lambda(0)=0$ to  $\lambda(\mathcal{T})= \lambda_0$, starting from an initial
state which is an angular momentum eigenstate $ \vert S=2,m\rangle $ (left, $\lambda_0 = 1$,  $ m=-1$;  right, $\lambda_0 = \lambda^{\star}(0)= 0.838$, $m=0$). The big black dot
stands for the  adiabatic prediction for $ \langle S_z \rangle  $. The red curves   are  for 
 a linear increase of  $\lambda(t )$, while the black ones correspond 
 to the OT choice, {\it i.e.}  
  $\dot{\lambda}(t )= f( \frac{t}{\mathcal{T} } )  /f(0)$, where  $f(s)$ is a Blackman pulse. The rather strong oscillating non-adiabatic   corrections to  $ \langle S_z \rangle  $  are clearly swept away by this latter  choice, provided $ \gamma_S \, B \, \mathcal{T} \geq 25 $. }
 \label{adiabla}
 \end{figure*}
In the previous subsections Berry's phases were induced by cyclic
variations of the periodic Euler angles performed in  an initial state
$\hat\psi(S m;\lambda_0 )$. In fact to prepare this eigenstate one  has to start from an eigenstate of $\vS$ and $S_z$ at a certain time t=0. 
During  the time
interval $0 \leq t\leq {\cal T}$,  the evolution of the spin  system  is governed
 by $ {\cal H}(\lambda (t)) = S_z /\hbar + i \lambda (t)( S_x/ \hbar)^2 $. 
 Writing   $ \lambda (t ) = \lambda_0  \, g( t/\cal T)  $, we assume that $ g(s)  $ 
  rises ``slowly'' from 0  to 1. 
 We  now ask asks   whether,  for a given values    of the rising time  $ \cal T$,   there might be  
  a better  choice for  $g(s$)   than just $ g(s)=s   $  when $ \lambda_0 \sim 1$,
   if one wishes to keep the non-adiabatic corrections below a predetermined  level .      
   \emph{In the present subsection, we have chosen ${  1/  (\gamma_S \, B ) }$  as the time unit: all the time  dependent physical quantities  and
their time derivatives  are   functions  of $\tau = t \, \gamma_S \, B.$ }
 \subsubsection{ The two-level spin system $ S=2,m=\pm 1$}
  To get some insight into this   problem, let us first consider the simple case of the
adiabatic evolution of the state $S=2,\, m=\pm 1$ governed by the
Hamiltonian ${\cal H}_2(t) = {\cal H}_{odd}(2 , \lambda(t)) $ defined by the 
$2\times 2 $ matrix of Eq.(\ref{H12}).  By performing the
change of variables $\zeta = \arctan{( \frac{3}{2} \lambda) }$ with $-
\pi/2  < \zeta  < \pi/2$, one can write ${\cal
H}_2(t) $ under the convenient form:
 \be
 {\cal H}_2(t) =  \sec {\zeta} ( \sigma _x\sin{\zeta}   + \sigma_z
\cos{\zeta})+
   \frac{5}{3} \tan (\zeta ) \sigma _0 ,
\ee
where $\sigma_0$ is the $2\times 2$ unit matrix and $\sigma _{x, y,z} $
are the standard Pauli matrices.
We define $\Phi(2 m; t)$ as the state vector obeying the Schr\"odinger
equation relative to ${\cal H}_2(t) $ and satisfying  the initial
condition $\Phi(2 m; 0) =  \vert 2,m\ket .$
We introduce the unitary rotation  matrix $V_2(\zeta) = \exp{-i \zeta
\sigma_y/2}$.   This is the rotating frame Hamiltonian $\wt {\cal H}_2(t)
$ which governs the evolution of $\wt \Phi(2 m; t) = V_2^{-1}(\zeta) \,
\Phi(2 m; t)$.
After some simple algebra we arrive at the familiar expressions:
  \bea  \widetilde{ \mathcal{H}}_2 (t)     &=
& \widehat{ \mathcal{H}}_2-i \, V_2  (-\zeta ) \frac{d}{d t}  \,
  V_2  (\zeta ) , \nonumber \\
    &=&\frac{5}{3} \tan (\zeta ) \sigma _0  +   \sec (\zeta ) \sigma _z   -
    \frac{1}{2} \;\dot{\zeta } \,  \sigma _y .
       \eea
       If $\zeta $ is assumed to grow linearly during the time interval $0
\leq \tau \leq  \gamma_S  B\,{\cal T}$,  $\widetilde{ \mathcal{H}}_2 (t)$  can be identified
with the Hamiltonian describing a square Ramsey pulse.
   The same  analysis can  be repeated  for the $  S=1, m=  \pm 1 $. The
result   looks very
      similar : the  spin dependent part of the rotating frame  Hamiltonian
      $ \widetilde{ \mathcal{H}}_1 (t) $ is  formally
      identical, but with one  significant  difference, namely $\tan \zeta =
  /2 $.

       In section VI, we shall study   the time evolution of a four-spin
system governed by
       the direct sum of  the two  Hamiltonians:
       $   {\widetilde{\mathcal H}}_{1}(t)  \oplus  {\widetilde{\mathcal
H}}_{2}(t) $
          relative to the same  value of $\lambda$.
          So  we will have to use  the two  different expressions of 
$\dot{\zeta } $,  for S=2,
      $\dot{\zeta_2 }  \rightarrow  6 \dot{\lambda } /(  9 \lambda
^2 +4) $, and for S=1, 
      $\dot{\zeta_1 }  \rightarrow 2 \dot{\lambda } /(\lambda ^2+4)$. 
   If  $\lambda[t] $  is rising linearly, the $\dot{\zeta}[t]  $ pulses  are replaced by trapezoidal Ramsey pulses with  sharp edges leading also  to
non-adiabatic oscillating corrections.

     There is  a standard way to wash them out which is used in nuclear 
     magnetic resonance and atomic interferometry experiments. This consists in using the so-called
Blackman  pulse shape \cite{Black}.
      Adapted to the present context, it  leads  the following choice of  
$\dot{\lambda}(t)$:
     \bea
      \dot{\lambda}_{BM}(t)&=& \lambda_0 f(\frac{t}{{\cal T}})/f(0),
\nonumber \\
     \hspace{-5mm}\text {where} \;\;  f(s) & =& 0.42  -0.5 \cos{(2\pi s)}
+ 0.08 \cos{(4\pi s)}. \label{BM}
     \eea
We have solved numerically the Shr\"odinger equation for $\lambda_0 = 1$ for the two
different choices, Blackman and trapezoidal  pulses for
$\dot{\lambda} (t)$. The important quantity for our purpose is  $
\langle S_z (\mathcal{T} )\rangle  $, since it governs the magnitude
of Berry's phase induced by the cyclic variation of the Euler-angles. It is
important to note that $ \langle S_z\rangle$ is affected to first-order in
$\dot{\lambda(t)}$ in contrast with the dynamical phase which is
modified only to second order.

      We have plotted $ \langle S_z (\mathcal{T} )\rangle  $ in Fig. \ref{adiabla} 
(left-hand curves), as function of $\gamma_S  B\,{\cal T}$.
 The big (black) dot on the ordinate axis shows the adiabatic
prediction.   The use of the Blackman pulse for $\dot{\lambda}
(t)$ leads to spectacular convergence towards the adiabatic limit: for $\gamma_s B\, {\cal T} \geq 25  $ the non-adiabatic correction to $ \langle S_z (\mathcal{T} ) \rangle  $ 
plunges down below the 0.02 $\%$ level. We have also computed the
phase shift, which is purely dynamical in the present case.  As
expected, the non-adiabatic corrections are much less sensitive to the
time dependence of $\lambda(t)$ and we find that in both cases  $
\vert  \phi_D(exact)-\phi_D(adiab) \vert < 0.008$  when $   \gamma_S B
\mathcal{T}  \geq 25 $.  Similar calculations for
$S=\vert m\vert=1$ and the initial value $m=-1$, for $\lambda_0= 1$ indicate that  corrections (not shown) are an order of magnitude smaller. 

\subsubsection{More than two levels}
 We  now  extend  the foregoing ``rotating'' frame  method 
 to non-adiabatic  processes involving the mixing of more  
  than two levels.   As noted in section II.A, the Hamiltonian  
    $  {\cal H}(\lambda (t)) = S_z /\hbar +  \lambda (t)( S_x/ \hbar)^2 $ 
        is described in the  angular momentum
 basis  $ \vert S \, m \ket $  by a real symmetric matrix. This    implies that its eigenstates   $ \hat \psi(S m; \lambda)$  can be written as real vectors. An evident
  consequence  is the following identity, obtained by taking the derivative
  of the normalization condition   of the  real eigenvectors:
  \be
  \bra \hat\psi(S m; \lambda ) \vert  \frac {\partial }{\partial \lambda
}\hat\psi(S m; \lambda) \ket   =  0.
\label{RV}
\ee
  It   implies  that the  adiabatic phase is purely dynamical.
   Let us introduce now  the  transformation  matrix:
    \be
    V_S ( \lambda )=\sum_{m} \vert\hat\psi(S m; \lambda )\ket \bra S m\vert .
    \label{VS}
    \ee
    One can directly verify that  $V_S ( \lambda )$ transforms $ {\cal H}( S m; \lambda)$ into a diagonal matrix within the basis $\vert S m  \ket $,   by writing
  $ 
  V_S ( \lambda) \vert S m_1 \ket = \sum_{m} \hat\psi(S m; \lambda )
 \bra S m\vert S m_1 \ket=\hat\psi(S m_1; \lambda ).   
 $
The time evolution in the ``rotating'' frame is  then governed by the following
Hamiltonian:
  \bea
 \widetilde{ \mathcal{H}}_S(t )& =  &V_S ^T \, {\cal H}( S m;
\lambda)\,V_S
 - i \dot{\lambda} \,  V_S ^T \frac {\partial } {\partial \lambda   }V_S ,
\nonumber \\
 &=& \widehat{ \mathcal{H}}_S(\lambda)+\Delta\widetilde{ \mathcal{H} }_S (t ).
  \eea
  The expression  
  $\widehat{ \mathcal{H}}_S(\lambda)= \sum_m \mathcal{E}( Sm;\lambda )
   \vert   S m  \ket\ \bra S m  \vert $ follows from eq. (\ref{VS}),
   while  the identity (\ref{RV})  implies that 
    $  \Delta\widetilde{ \mathcal{H} }_S (t ) $  is non-diagonal:
 \bea
  \Delta\widetilde{ \mathcal{H} }_S (t )= 
     \sum_{m_1 \neq m_2}    \vert   S m _2    \ket\    \bra S m _1  \vert 
     \times \nonumber\\
 -i \dot {\lambda}\,       \bra \hat\psi(S m_1; \lambda ) \vert 
\frac {\partial } {\partial \lambda   }\hat\psi(S m_2; \lambda ) \ket
\,.   \nonumber    
\eea    
By  using  an identity obtained from the  time  derivative of  the eigenvalue equation (\ref{calH}),  
we can  rewrite the r.h.s of the above equation as:
     \bea
     \Delta\widetilde{ \mathcal{H} }_S (t )&=&   -i \dot {\lambda}(t)\times
     \sum_{m_1 \neq m_2} \frac{     \vert   S m _2    \ket\    \bra S m _1
 \vert }
     { \mathcal{E}( S m_1; \lambda)- \mathcal{E}( S m_2; \lambda) } \times
 \nonumber  \\
     & &  \bra \hat\psi(S m_1; \lambda ) \vert( S_x/\hbar)^2 
 \vert \hat\psi(S m_2; \lambda ) \ket \,.
     \eea
   The above formula can be considered as an adaptation of a more general one
to be found  in
 \cite{Berry2} and  references therein. 
  One must stress that energy denominators  $  { \mathcal{E}( S m_1; \lambda)- \mathcal{E}( S m_2; \lambda) }$ 
   involve levels  for which $ m_1-m_2$ is even integer.
For instance, in the cases $S=2, \,  m= 0,\pm 2 $  the two energy differences involved are varying  
slowly within the interval $ 0\leq \lambda \leq \lambda_0 = 0.838. $ 
 It follows that the time dependence
of $ \Delta\widetilde{ \mathcal{H} }_S (t )$ is   
 dominated by $ -i\, \dot {\lambda}(t)\,$. 
Let us assume  that  $\lambda(t) $  is increasing  linearly  within the time interval
$  0 \leq  t\leq \mathcal{T} \, :\,  \lambda(t) =\lambda_0  t / T$.
The effect of  $ \Delta\widetilde{ \mathcal{H} }_S (t )$, as in the previous cases,
is equivalent  to a r.f.   trapezoidal pulse with sharp edges.   
     In the right-hand curve of Fig.5, we have plotted the average polarization as a function
of the rising time $ \mathcal{T}$ for $ \lambda(t) $, towards the value $
\lambda_0=\lambda^*(0) =0.838$. 
  There is a a very rapid damping of the oscillating non-adabatic
corrections, passing below the level  of  $ 0.2 \% $  for  $ \gamma_S
\, \mathcal{T} \geq 25 $ when $\dot {\lambda}(t) $ has
 a Blackman pulse shape. This is in strong contrast with  the case of a 
linearly rising  $ \lambda(t) $
 where, for $ \gamma_S \,B \mathcal{T} \geq  25$,  the non adiabatic 
oscillating corrections   have an 
 amplitude  of  about $ 10\%$, decreasing only very slowly for longer
rising times.
 The conspicuous   chaotic behaviour in the 3-level case
 reflects the fact that the two frequencies involved
 have in general an irrational ratio.
   \subsection{Concluding remarks}
 Let us sum up the results of this last section.
  We have shown that if $ \lambda(t)$ is a linear 
 function within the finite time  interval $ 0 < t < T $, vanishing elsewhere,
 oscillating non-adiabatic corrections are generated by the sharp jumps of 
 $\dot{\lambda}(t) $ at $t=0$ and $t=T$. We have found that the remedy is to choose for $\dot{\lambda}(t) $ a Blackman pulse shape. It is clear that one should 
 use the same prescription for $ \dot{\alpha}(t) , \dot{\varphi}(t)$,  
 for taming analogous unwanted oscillations.   By solving exactly the Schr\"odinger  equation governed by the Hamiltonian  $\wt H_{//}(t)$  of Eq. (\ref{Htdparal}), as illustrated in Sec. VI and in a separate paper \cite{bou3} on two definite examples, we have indeed found that this remedy works beautifully.      
 
 It should be stressed that
   the near-degeneracies of opposite-parity level  pairs
  $ (  {\cal E}(m, \lambda),{\cal E}(m-1, \lambda) )$, which occurs when 
  $ m>0 $ and  $ \lambda > 0$, do not affect the validity of the adiabatic 
approximation  for Berry's cycles if  $ \alpha(t) $ and $ \lambda (t)$
are the only time-dependent parameters. This simplification is a consequence of the 
symmetry properties of the Hamiltonian  $H(\vB(t),\vE(t))$ resulting from the $\vE$ and $\vB$ orthogonality.  

Furthermore,   choosing for  $ \lambda (t)$ the 
  ``magic" expression $ \lambda^* (\frac{ \dot{\alpha} (t) }{\gamma_S \, B} )$ 
given by   Eq.(\ref{lamagfit})
 makes  the  difference of the Berry phases relative to the  mirror cycles   
  $ ( \alpha(t) ,\lambda (t) )$ and $ ( -\alpha(t) ,\lambda (t) )$   free of 
    non-adiabatic corrections:    
all the correcting terms   $ \propto   ( \dot{\alpha} (t)/\gamma_S \, B )^{ 2 q+1} $ vanish  
  whatever $q >0 $. 
                 \section {Holonomic entanglement of ${\mathbf{N}} $ non-correlated 1/2 spins} 
\subsection{Introduction}
The purpose of this section  is to show that  Berry's cycles studied is
this paper
are able to entangle  $N$  non-correlated one-half spin states, while such
an effect would  not be present with  linear spin couplings.  Since the 
analysis presented here  is exploratory, we shall  concentrate upon  the 
$N=4$ case  which  involves already the   basic features  of our
entanglement procedure.

   It is  convenient  to classify the set of  the four  states  with
respect to the eingenvalues
   of $ S_z=\sum_i  s_{z i }$: $ M= \sum_{i=1}^{4} m_i $. The case $ M=2$
is trivial. The case
$ M=1$ is the most interesting one for our purpose and will be the main 
subject of this  paper.
The case $ M=0$ will be discussed briefly at the end of this section. The
results for
 negative values of $M$  are readily obtained from the positive ones by using
 the reflection laws introduced in section III.
 The  vector space generated by the linear combinations of non-correlated
spin  states with $ M=1$  has a dimension  four  and  the natural choice 
for a basis is the set of the four orthogonal  states:
   \bea
    \Phi ^{(1)}  & =&
    \vert -\frac{1}{2} \ket \otimes  \vert \frac{1}{2} \ket \otimes
    \vert \frac{1}{2} \ket   \otimes  \vert \frac{1}{2} \ket,\nonumber\\
  \Phi ^{(2)}& =& \vert \frac{1}{2} \ket \otimes  \vert -\frac{1}{2} \ket
\otimes
  \vert \frac{1}{2} \ket  \otimes  \vert \frac{1}{2} \ket, \nonumber\\
 \Phi ^{(3)}& =& \vert \frac{1}{2} \ket \otimes  \vert \frac{1}{2} \ket
\otimes
 \vert -\frac{1}{2} \ket \otimes  \vert \frac{1}{2} \ket, \nonumber\\
    \Phi ^{(4)}&=& \vert \frac{1}{2} \ket \otimes  \vert \frac{1}{2} \ket
\otimes
     \vert \frac{1}{2} \ket \otimes  \vert -\frac{1}{2} \ket .
    \label{4Phi}
    \eea
 
     We shall assume that the time evolution of the four-spin state $
\Psi_4(t) $ is
    governed  by  the time-dependent
   Hamiltonian $ H_N (t) = H( \vB(t),\vE(t) ) $, which looks formally
   like  the quadratic spin Hamiltonian
   of Eq.(\ref{themodel})
   discussed extensively in the present paper:
  \bea
   H(\vB(t),\vE(t))&=&   U(R(t))\, \hat{H}(B(t),E(t))  \,  U ^{ -1}(R(t))
,\nonumber\\
    \hat{H}(B(t),E(t)) &=& \gamma_S \, B(t) ( S_z/\hbar + \lambda(t) (S_x/\hbar^)2 ).
\nonumber
  \eea
 \emph {The crucial difference lies in the fact
   that  $\vS$  is meant to  be  the total spin operator
     $\vS = \sum_{i=1}^{N} \vs_i $. }

     We shall take as initial state for $\Psi_4(t) $, one of the
non-corellated
     states appearing in equation (\ref{4Phi}), say, $\Psi_4(0)= \Phi
^{(1)}$.
          By expanding  $ H_N(t)$ in terms of single-spin operators
     one gets a sum of   spin-spin interactions:
      $ \propto \vs_i\cdot \hat n \;\;\vs_j \cdot \hat n$.
     where  $\hat n$    is given by $ \hat n= U(R(t)) \hat x$.
    This indicates  that $ H_N(t)$  can create spin correlations in  
the state  $\Psi_4(t) $.  As a final remark, we stress
that, by construction,   $ H_4 (t) $ is invariant under all permutations  of
the  $N=4$ spins. This invariance property will play an   essential role in
 our entanglement procedure.

\subsection{Expansion of  a  non-correlated four spin state into a sum  of the
$ \vS^2$ eigenstates  }
 As a first step, we shall  construct four angular momentum eigenstates
which are linear combinations of the four  states $\Phi ^{(i)}$.
 Using the rules of adddition of quantum angular momenta, one finds that
the possible eigenvalues
of   $ {\vS}^2 $ ,   $\hbar^2 S(S+1) $,   correspond to   $S=2$ and  $S=1$.
 There is a unique way to construct
the state $ S=2 $.
  It is obtained by applying the operator $ S_{-}= S_x - i S_y$  upon the
state $S=2,M=2 $ {\it i.e.}
$ \Psi_{2\;2}= \vert \frac{1}{2} \ket \otimes  \vert \frac{1}{2} \ket \otimes
    \vert \frac{1}{2} \ket   \otimes  \vert \frac{1}{2} \ket $. One gets
immediately:
\be
\Psi_{2\,1}=\frac{1}{2} ( \Phi ^{(1)} +\Phi ^{(2)}+\Phi ^{(3)} +\Phi
^{(4)}  ).
\label{Psi2_1}
\ee
The above  state is clearly invariant under all permuations of the 4-spin
states.

One must now  construct  three orthogonal   states  with $ S=M=1, 
\Psi^{i}_{\,1\;1} $
which will have different symmetry under the  permutations of the four spins.
If one ignores the orthogonal condition, it is  easy  to get three $ S=1$
states which are linearly independent. The method consists in taking a
linear combination  of 2 states $ \Phi^{(i)}$
in such a way as to  factor out  the  two spin singlet  state:
$ (\vert \frac{1}{2} \ket \otimes  \vert -\frac{1}{2}\ket  -\vert
-\frac{1}{2} \ket \otimes  \vert \frac{1}{2} \ket )/\sqrt{2} $, which is
invariant  under the rotation group.
 Let us give a typical example of such a construction:
\bea
\Phi^{(1,2)} &=& (\Phi ^{(1)} - \Phi ^{(2)})/\sqrt{2}, \nonumber \\
&=&\frac{1}{\sqrt{2}} \((\vert \frac{1}{2} \ket \otimes
  \vert -\frac{1}{2}\ket  -\vert -\frac{1}{2} \ket \otimes   \vert
\frac{1}{2} \ket \))\otimes \vert \frac{1}{2} \ket   \otimes  \vert
\frac{1}{2} \ket \nonumber\\
  \label{Phi12}
   \eea
 is a good candidate for a $S=M=1$ state.
Two similar $S=M=1$ states   $\Phi^{(1,3)} ,\Phi^{(1,4)} $ are obtained 
from $\Phi^{(1,2)}$
by performing the cyclic permutation $(234)$.
 These three states are linearly independent but  a  basis  with
 orthogonal  states would  be more convenient.
 Since symmetry will be the basic tool in this subsection,
    let us start with   the more symmetric $ S=M=1$  state :
  $ \frac{1}{2}( \Phi ^{(1)} - \Phi ^{(2)}+\Phi ^{(3)} - \Phi ^{(4)})$.
  The  two other states are  obtained  by applying  the cyclic permutation
(234). One arrives then to   the following  basis for the  $S=M=1$
subspace, easily seen to be orthogonal:
  \bea
  \Psi^{1}_{\,1\;1}&=
  & \frac{1} {2} ( \Phi ^{(1)} - \Phi ^{(2)}+\Phi ^{(3)} - \Phi ^{(4)}),
\nonumber \\
   \Psi^{2}_{\,1\;1}&=
  & \frac{1} {2} ( \Phi ^{(1)} - \Phi ^{(3)}+\Phi ^{(4)} - \Phi ^{(2)}),
\nonumber  \\
    \Psi^{3}_{\,1\;1}&=
  & \frac{1} {2} ( \Phi ^{(1)} - \Phi ^{(4)}+\Phi ^{(2)} - \Phi ^{(3)}).
  \label{Psi1_1}
  \eea
   The three  above states have, different symmetry properties under
  the  three permutations (23),  (24), (34), which are listed below:
  \bea
  \{  (23 ) , (24), (34) \} \Psi^{1}_{\,1\;1}&= &\{-1,\,1,\,-1\}
\Psi^{1}_{\,1\;1},
  \nonumber \\
 \{  (23 ) , (24), (34) \} \Psi^{2}_{\,1\;1}&=& \{\;1,\,-1,\;1\;\}
\Psi^{2}_{\,1\;1},
 \nonumber \\
 \{  (23 ) , (24), (34) \} \Psi^{3}_{\,1\;1}&=& \{-1, -1,\,1\,\}
\Psi^{3}_{\,1\;1}.
 \nonumber
 \eea
 The four eigenstates $\Psi^{i}_{\,1\;1}$  and $\Psi_{\,2\;1}$ 
constitute a complete
ortho-normal  basis for the four spin one-half states with $ M=1$. The
expansion of the
 four  non-corrrelated states $ \Phi^{(i)}$ are readily obtained from
equations
 (\ref{4Phi}) and (\ref{Psi2_1})
\bea
\Phi ^{(i)}&=&\frac{1}{2} \(( \sum_{j=1}^{j=3}  a_{i\,j}
\,\Psi^{j}_{\,1\;1}+ \Psi_{2\,1}\)),\;\; {\rm with }
    \label{PhivsPsi}     \\
   a_{1\,j} & =&\{ 1,1,1\},\; \; \; \;\; \;a_{2\,j}=\{-1,-1,1\},  \nonumber \\
    a_{3\,j}&= &\{1,-1,-1\} ,\,  a_{4\,j}= \{-1,1,-1\}.
   \eea

\subsection{ The adiabatic evolution of  four non-correlated,
 1/2 spin states governed  by $ H( \vB(t),\vE(t) )$  }

  The Hamiltonian   $ H_4 (t)=H( \vB(t),\vE(t) ) $, as noted before, is
invariant under either one of the three ``parities''
       $\epsilon_{(23)},\epsilon_{(24)}, \epsilon_{(34)} $,  associated
with the permutations  (23),(24) and (34).  
   As a  consequence, the matrix elements:
   $ \bra \Psi^{i}_{\,1\;M} \vert  H_4 (t) \vert \Psi^{j}_{\,1\;M^{\prime}}\ket $
for $M=M^{\prime}=1$ vanish if $ i \neq j $ . Moreover, the same
result  holds  if  $M=\pm1$, 
 $ M'=\pm1$ and $M\neq M^{\prime}$,
  since   the lowering operator  $ S_x-iS_y$ commutes with all the four
spin permutations.
 Using  the same symmetry arguments,  one sees immediately
  that the matrix elements
   $ \bra \Psi_{\,2\;M} \vert  H_4 (t) \vert \Psi^{i}_{\,1\;M'}\ket $ vanish
   whatever the values of $ i, \,M, \, M' $.
   If $ U_4(t) $ stands for the unitary
  operator  associated   with the quantum evolution governed by    $ H_4
(t) $,  the four states
  $ U_4(t) \Psi_{2\;M}$,  and and $U_4(t)\Psi^{i}_{1\; M'} $ will have the same 
permutation  symmetries
  as their parent states. In conclusion,
   \emph{the four states $ \Psi_{2\,M} ,  \Psi^{i}_{1\,M'}$
   behave vis \`a vis  the Hamiltonian $H_4(t)$, as if they were associated
    with  isolated spins  S} and we can apply to them the
results derived in the previous sections.

Our   Berry's cycle is organized in three steps. At $t$=0, $H_4(0) = 
\gamma_S \, B(t)  S_z $. 
 The first step $ 0\leq t \leq {\cal T}$  involves
 the adiabatic  ramping of $\lambda(t)$ from 0 to $ \lambda_0 \simeq -1 $.
In the second step ${\cal T} \leq t  \leq 3 {\cal T}$, one proceeds to the
rotation of   $\vE $
around $\vB$ by an angle 3$\pi$, while keeping $\lambda(t)= \lambda_0$.  For the third step, $3{\cal T}\leq t
\leq  4{\cal T}$,    $\lambda(t)$   makes an adiabatic return
   to its initial  value $\lambda=0 $.
  During the whole cycle  the time dependences of    $\dot{\lambda}(t)$
and $\dot{\alpha}(t)$  are described by  Blackman functions. The effect
of this choice is  to tame
the non-adiabatic oscillating corrections which would be generated, if
 these parameters had discontinuous time-derivatives.

  We have performed an exact  theoretical  analysis of the cycle by
solving the corresponding Schr\"odinger equation in the rotating frame
with $\gamma_s B {\cal T}=25$.  We have found that the adiabatic
approximation is working to better than 0.1$\%$.
 It turns out that somewhat accidentally the difference of the dynamical
phases for $S=2, M=1$ and $S=1, M=1$ is close to 0 {\it modulo} 2$\pi$.  
Slightly tuning the $\lambda$-rising (-lowering) times can improve the
cancellation to the 0.1$\%$ level.
        The cyclic evolution of the   four spins, initially
non-correlated, appears then as completely governed by the
Berry's phases $\beta(2,1; \lambda_0)$
and  $\beta(1,1; \lambda_0) $.
 \subsection{ The final construction of the  four-spin holonomic
entangled state }
At the end of the Berry's  cycle, $\Phi ^{(i)}$ is transformed into the following state:
      \bea
   \Phi ^{(i)}_{BP}(\lambda_0)&=& \frac{1}{2} \{ \exp( i  \,\beta(
2\,1;\lambda_0 ) )\,  \Psi_{2\,1} +\nonumber\\
        &&   \exp(\, i \, \beta( 1\,1;\lambda_0 ) )  \sum_{j=1}^{j=3} 
a_{i\,j} \,\Psi^{j}_{\,1\;1}   \}.
   \eea
 Using Eq.(\ref{PhivsPsi})  to express the sum $ \sum_{j=1}^{j=3}
 a_{i\,j} \,\Psi^{j}_{\,1\;1} $  in terms of  $ \Phi
^{(i)} $
   and $  \Psi_{2\,1}$,  one obtains the final expression:
   \bea
    \Phi ^{(1)}_{BP}( \lambda_0)&=&
   \exp\(( \, i \,  \beta( 1\,1;\lambda_0 ) \))\(( \Phi ^{(1)}- 
\frac{1}{2} \Psi_{2\,1}  \)) + \nonumber \\
    &&  \frac{1}{2}    \exp\(( \, i \,\beta( 2\,1;\lambda_0 ) \))
\Psi_{2\,1}.
   \eea
(We can verify on this expression that if $ \Delta \beta (\lambda_0 ) = \beta( 2\,1;\lambda_0 ) -
\beta(1\,1;\lambda_0 )=0$, then $\Phi ^{(1)}_{BP}$ coincides with $ \Phi
^{(1)}$ up to a phase, as  one expects).

  From the energy curves given in Fig.1 one sees  that the region
   $ \lambda > 0  $ should be, in principle avoided, since the nearly
crossing   levels  $ S= 2, M=2$ and $ S=2, M=1$ could  spoil the
validity of the adiabatic approximation, while the region $\lambda <0$ 
is much more favourable. However,   in the present
   context  where $ \lambda$ and $ \alpha $ are the only time-dependent
   parameters, this  is only  a protection against stray magnetic fields
orthogonal to the main field, since $H_4(t)$ has no $\Delta M = \pm 1$ 
matrix elements.
 The    two Berry's phases   are given explicitly  by rather simple
expressions:
    \bea
      \beta( 2\,1;\lambda_0)  & =&   3 \pi  \left(\frac{2}{\sqrt{9
\lambda_0 ^2+4}}-1\right), \nonumber \\
  \beta(1\,1;\lambda_0)& = & 3 \pi  \left(\frac{2}{\sqrt{\lambda_0
^2+4}}-1\right).
     \eea
     
     In order to obtain the maximum  entanglement  of    $\Phi ^{(i)}_{BP} $ 
after a rotation of $3\pi$, it is interesting  to choose $
\lambda_0=\lambda_{max} =-0.97 $ which leads to  $  \Delta \beta
(\lambda_{max} ) =-\pi $.
              Replacing  $\Psi_{2\,1} $    by its expression in terms of $
\Phi^{(i)}$
              ( Eq.(\ref{Psi2_1}))
            we arrive  at the   final   expression of
     the four   quantum   entangled states,  generated by the Berry's
cycles with
     $ \Delta \beta (\lambda_0 ) =  -\pi$,
        from any of the four non-correlated  states   with  M=1  listed in
 equation ( \ref{4Phi} )
         \be
      \Phi ^{i}_{BP}(\lambda_{max} )  =  \exp (i \chi    )
       \(( \Phi ^{(i)}- \frac{1}{2} \sum_{j=1}^{4 }\, \, \Phi^{(j)}\)),
             \ee
       where     $\chi= \beta( 1\,1;\lambda_{max})$.
        In order to make more apparent the holonomic entanglement resulting 
       from the Berry cycle, let us rewrite
        $\Phi ^{1}_{BP} (\lambda_{max} )$, using the explicit forms of 
         the non-correlated states $ \Phi ^{(i)}$ given in Eq. (\ref{4Phi}):
        \bea 
      \Phi ^{(1)}_{BP}(\lambda_{max} )& = & 
     \frac{\exp (i \chi    )}{2} \,
      (  \, \vert -\frac{1}{2}\ket \otimes  \vert \frac{1}{2} \ket \otimes  
    \vert \frac{1}{2} \ket   \otimes  \vert \frac{1}{2} \ket \nonumberÊ\\
  &&  -   \vert \frac{1}{2} \ket \otimes  \vert -\frac{1}{2} \ket \otimes  
  \vert \frac{1}{2} \ket  \otimes  \vert \frac{1}{2} \ket \nonumber\\
 &&- \vert \frac{1}{2} \ket \otimes  \vert \frac{1}{2} \ket \otimes  
 \vert -\frac{1}{2} \ket \otimes  \vert \frac{1}{2} \ket \nonumber \\
 && - \vert \frac{1}{2} \ket \otimes  \vert \frac{1}{2} \ket \otimes 
     \vert \frac{1}{2} \ket \otimes  \vert -\frac{1}{2} \ket  \,).
      \eea

     A similar approach can be used to treat the case of four 
non-correlated spins with
      $ M= \sum_{i=1}^{4} m_i =0$.  Among  the three angular momentum
      states $ S=2, 1,0;M=0$, only the state $S=2, M=0$ has a non-vanishing
      Berry's Phase. As a consequence,   one gets  an entanglement which
is weaker than
     in   the above  $ M=1$ case where it is maximum.
     The case of 8 one-half spins (or 8 Qbits) with $ M=1 $  is of
particular interest since it  involves four   non-vanishing Berry's Phases for $
S=1,2,3,4$ but it will require more powerful  Group Theory tools, like the Young
Operators \cite{MHam}. However,  the basic ideas of  the entanglement 
procedure  would   remain the same.

  The reader might have the impression that the above method  of entanglement
       is constrained  by the fact that Berry's phases depend upon few
parameters.
    However one should keep  in mind that $\beta (S M;\lambda) $ is
actually given   by a Bohm-Aharonov integral involving  an Abelian
gauge field along a closed loop drawn upon    the four-dimension
manifold $ \vC P^2 $, which implies a considerable  freedom (see
Eq. (\ref{betageom})).

                 From a practical point of view entanglement methods based on
geometric phases are known to present advantages of
robustness: they are resilient to certain types of errors in
the control of the parameters driving the quantum evolution
\cite{Sjo}.


    We end this section by giving an example where the collective
spin-spin interaction between a few spins, similar to that described
by $H_N(t)$, has been successively implemented. This concerns the case
of two or four alkali-like ions, without nuclear spin, in an ion-trap.
The idea is to realize  illumination of the N ions (N Qbits) with two
lasers fields having  two different frequencies so that the two-photon
process, exciting any pair of ions in the trap is resonant, but neither of
the frequencies are resonant with single excitation of an ion \cite{molm,
sore}. Realization performed for two and four ions in a trap \cite{sack},
can be generalized to more ions.  In the present context it would be
necessary to match the magnitude of the dc magnetic field and the
radiation fields. According to the analysis given above,  rotating the
linear polarization of the radiation fields around the $\vB$-field should
make it possible to generate holonomic entanglement of the Qbits.
 \section{Summary and perspectives}
 \subsection{Synopsis of the paper} 
 The purpose of this paper is the theoretical study of the Berry's phases, generated in cyclic evolutions of isolated spins of arbitrary values. The spins are assumed to be non-linearly coupled to time-dependent
external electromagnetic fields (possibly effective) via the
superposition of a dipole and a quadrupole  couplings.
 Configurations leading to degenerate instantaneous  eigenvalues are avoided.
In other words non-Abelian Berry's phases are not considered.
We also assume that the two  effective fields
 are orthogonal, a mild restriction but with many advantages. This implies
 several discrete symmetries of the spin Hamiltonian which simplify
considerably
 the algebra. For instance, two angular momentum states having different
 quantum numbers  $ m_1$ and $ m_2$ are coupled if, and only if,
$ m_1-m_2$
  is an even integer. Furthermore, for  $ S=1$,  the geometric space of
the Hamiltonian
  parameters is isomorphic to the density matrix space  $\vC P^2$  which
makes the Berry's and   Aharonov-Anandan  geometric phases
mathematically identical.

Using rotation group theory, and the aforementioned discrete symmetries,
we obtain compact  expressions for the instantaneous
eigenfunctions of the Hamiltonian, labelled by the magnetic
quantum number $m$, valid in the limit of small  quadrupole coupling.

 We derive  an  explicit compact expression  for the Berry's phase
 $ \beta(m) $ in terms of the usual Euler angles $ \varphi, \theta, \alpha
$  associated
with the trihedron defined by  the $\vB , \vE$ directions, and an extra
dimensionless parameter $ \lambda $ giving the quadrupole to dipole
coupling ratio. The result is given as a loop integral of an Abelian field
along a closed circuit drawn upon the parameter space $\vC P^2 $,
$$
  \beta(m)=     \oint_{{\cal C }}   p( m, \lambda) ( \cos \theta \,
d \varphi +  d \alpha )
 - m\,(d \varphi +d \alpha )  .
$$
  where $\hbar\, p( m, \lambda) $, the average value of the spin angular
momentum
taken along the $ \vB $ field, is just the gradient  of the
eigenenergy  with respect to $ \vB$. The apparent simplicity
of the above formula  conceals the  geometry contained in $p(m, \lambda)$.
This important   feature  is   clearly exhibited in Berry's cycles
  where $ \lambda $ and $ \alpha $ are the sole varying parameters.
  The   case  $ S=2, m=0$, where the Berry's Phase is  vanishing
   for  a linear spin  Hamiltonian, is particularly spectacular:
$ p (0,\lambda)$ is an odd function of  $ \lambda $ and takes the value 1
when  $ \lambda=1$. This peculiar geometry  is best illustrated
in  Fig. \ref{geom} where  $ A_{\alpha }=   p(0,\lambda) $ is plotted
 against  the spherical   coordinate  $\tilde{\theta }$ defined by
    $\lambda=-2 \cot \tilde{\theta }$  with $  0<  \tilde{\theta } < \pi $.
   
  The non adiabatic corrections associated with  the Euler angles
 derivatives have been analyzed within the rotating frame
attached to the time-varying fields. The Coriolis effect generates an
extra magnetic field $ \Delta \vB$ which involves a linear combination
of the Euler-angle time derivatives. The longitudinal
component along the $ \vB $ field is the only one which survives when
$\alpha$  is the
sole time-dependent Euler angle. The corresponding Hamiltonian $
\wt{H}_{//}$ is devoid of any geometry. As a consequence, the phase shift
acquired  at the end of
 the cycle $ \tilde{\phi}(\vB + \Delta \vB)$ is purely dynamical.
  The Berry's phase  is  incorporated into the dynamical phase under the
form of its first-order contribution with respect to $ \eta= - \Delta
B_{//} /B =
  (\cos {\theta} \; \dot{ \varphi} + \dot{\alpha} )/(\gamma_S B )$.
 The higher-order terms give all
 the non-adiabatic corrections associated with $  \dot{\alpha} $ when it
is the only
 varying periodic parameter. We have also shown that the subset of these
corrections,
 odd under a reversal of $ \eta $, cancel exactly for    ``magic" values
$ \lambda =\lambda^*(\eta)$. This cancellation is implemented in the
experimental
project described in reference \cite{bou3}. The case of the non-adiabatic
corrections induced by
the transverse magnetic field $\Delta B_{\perp} /B$ is somewhat more
involved since it introduces a non-trivial geometry and, as a consequence,
a Berry's phase contribution
to be added to the one coming from the transverse dynamical phase.
An explicit evaluation of the complete lowest-order  total correction $
\Delta \beta_{\perp}/\beta $     is found to be   $ \propto(\dot{ \varphi
}/\gamma_S B )^2 $ up to a numerical coefficient computed explicitly for $
S=2, \,m=0,$ and -1. The results are displayed
in Fig. \ref{coradiab}.

This discussion hinges upon the implicit assumption that the finite value of
 $ \lambda $ is reached   by an adiabatic ramping governed by
 $ \mathcal{H}(  \lambda(t) )= S_z/\hbar+ \lambda(t) ( S_x/\hbar )^2 $, 
leading  to a pure eigen-state
 $\hat{\psi}(\lambda) $,  free of any non-adiabatic pollution. To analyse
this problem,
 it was  convenient to perform the time dependent unitary transformation $
V(\lambda(t))$ which  transform  $ \mathcal{H}(\lambda(t)) $,
into a diagonal matrix  within
the angular momentum basis. It follows from  the symmetry properties of $
\mathcal{H}(  \lambda(t) )$  that $ V(\lambda(t))$ can be taken as {\it
real and symmetric}. The Hamiltonian, which  governs
  the spin evolution in the transformed basis,    contains the  non-diagonal
   term $ i \dot{\lambda}(t)  V^T(\lambda(t))\frac {\partial } {\partial
\lambda} V(\lambda(t)) $.
   Since its  time dependence is dominated
   by the time derivative $\dot{\lambda}(t)$,   a linear increase of
$\lambda(t) $  would be equivalent  to
a rf pulse with sharp edges  leading to the oscillating non-adiabatic
corrections
 exhibited in Fig.\ref{adiabla}. 
  The  standard procedure to smooth them
  out is to use for $\dot \lambda(t)$ a Blackman pulse shape \cite{Black}.
  The  efficiency of the procedure for taming the non-adiabatic
oscillations is  clearly exhibited in Fig.\ref{adiabla}.
 There is a second assumption  implicit in the rotating frame analysis,
namely that the adiabatic  approximation be valid for the rotating frame Hamiltonian
$\wt{H}(t)$.
  By solving numerically the Shr\"odinger equation describing the whole
cycle, we have verified (see ref. \cite{bou3} for explicit numerical
results)   that,  indeed,   the adiabatic approximation works
beautifully provided one also uses a Blackman-pulse shape for the
angular speed, $ \dot{\alpha}(t) $.

In the last section,  we propose a procedure  to entangle a
system of N non-correlated  one-half spins (or N Qbits). It  involves   Berry's cycles
generated by an  Hamiltonian formally identical to one 
given in equation (\ref{themodel}), but with an important change:
$\vS$ stands now  for the total spin operator  $\vS = \sum_{i=1}^{N} \vs_i $.
  We have used a method which   is well  adapted, to the simple configuration of four
one-half spins. In view of  possible extensions  to $N$ spins with 
$N > 4$, we would like to rephrase our approach within a more general
framework. A configuration of $N$ non-correlated $\frac{1}{2}$-spins can be
described by a factorizable spin tensor of order $N$. It constitutes a
reducible representation of the $S(U_2)$ group.
 There is a general  procedure to decompose this tensor into irreducible tensors,
 associated with eigenstates of $\vS^2$ and $ S_z$. 
 With the exception of the trivial case $N=2$,  the angular momentum state $\Psi_{S\,
M}$ can  appear several times in the decomposition. According to a known
theorem of  Group Theory \cite{MHam}, all the states $\Psi_{S\, M}^i$ have 
different  symmetry properties under the permutation of the $N$ spins and
they can be organized into an orthogonal basis,
$\bra \Psi_{S\, M}^i \vert \Psi_{S^{\prime}\, M^{\prime}}^j \ket =
\delta_{S,S^ {\prime} } \delta_{M,M^ {\prime} }\delta_{i,j} $. 
  Let us now  consider the set of Hamiltonians  ${\cal{H}}(\vS) $
which are given by a  non-linear series expansion with respect to the spin
components $S_{x,y,z}$. By construction,  ${\cal{H}}(\vS) $ is 
invariant upon any permutation of the $N$ spins. As a consequence, 
 ${\cal{H}}(\vS) $ is diagonal within the basis  $\Psi_{S\, M}^i$:
$
\bra \Psi_{S\, M}^i \vert {\cal{H}}(\vS, t) \vert \Psi_{S^{\prime}\,
M^{\prime}}^j \ket = \delta_{S,S^ {\prime} } \delta_{i,j}
\bra \Psi_{S\, M}^{i }\vert {\cal{H}}(\vS, t) \vert \Psi_{S,M'}^{i}\ket.
$
The above equation expresses  the  simple fact that ${\cal{H}}(\vS) $ is
acting upon the states $\Psi_{S\, M}^i$ as if they were  isolated spins $S$.
Any initial state of $N$ non-correlated spins can then  be written as
$\vert\Psi_N(t=0)\ket= \sum _{i\,, S} a_{S\,M}^i \vert \Psi_{S\, M}^i\ket. $
One performs now  an adiabatic  Berry's cycle along a closed loop
in the parameter space such that
$ {\cal{H}}(\vS, 0) = {\cal{H}}(\vS, T) = \gamma_S \vS \cdot \vB $.
At the end of the cycle the $N$ spin system is given by the following state vector:
$
\vert\Psi_N(T) \ket = \sum_{i ,\, S}  a_{S\,M}^i \exp{ i \beta (S \, M)}\vert \Psi_{S,\, M}^i \ket.
$
 
 With a suitable choice of the Berry's cycle, we have shown in the
particular case $N=4, M= 1$ that the final state is endowed with a maximal 
entanglement.
 Thus extension to higher values of $N$  is worth pursuing,  remembering
 that  according to quantum computing experts:
 ``entanglement, as with most good things, it is best consumed in
moderation'' \cite{gros}.
\subsection{Experimental perspectives}
In a separate work, we have explored the possibility of cold alkali atoms in their ground state to measure Berry's phase with atomic interferometers.
 The spin operator is then  identified with the total angular momentum $ \vF =\vs + \vI $. 
 The ac-Stark shift induced by a linearly polarized light beam tuned off-resonance of 
 one resonance line,  
can induce the quadratic interaction if one accepts a few experimental compromises.  
The candidate for our
spin system is the ground state hyperfine (hf) level  $ 5S_{1/2},\, F=2 $ of  $^{87}$Rb.
 We  have found  judicious to tune  
the laser frequency  midway  between the two lines $  5S_{1/2},\, F=2 \rightarrow  5P_{1/2},\, F=1$, and $F=2$ . 
 The effective quadrupole coupling takes then  the simple form:
$H_Q =\frac{ \hbar \Omega^2 }{ \Delta{\mathcal{W}}_P }  (\vF\cdot \hat e)^2,$
where $\Delta{\mathcal{W}}_P $ is the $ 5P_{1/2}$ hf  splitting,
$ \Omega$ the Rabi frequency relative to the transition
$ 5S_{1/2} \rightarrow  5P_{1/2}$ and $\hat e $ the light polarization.  
 However there is a certain price  to be paid arising from  the instability of the
``dressed" ground state depicted by its decay rate 
$\Gamma_{dec}=\frac{4 \Omega^2}{\Delta{\mathcal{W}}_P^2} 
\Gamma_{5P_{1/2}} $ (where $\Gamma_{5P_{1/2}}$ is the spontaneous decay rate of the excited state).  
Although, with our choice of detunings, this effect does not affect the Berry phase value, 
it  can perturb seriously  the measurement process.  One is  clearly  facing a difficult  optimization
 problem, if one wants also  to keep  the non-adiabatic correction below the $0.1 \% $ level.
 We give here some  features of the solution described in ref. \cite{bou3}.
With an  external magnetic field  of 1 mG,  one can  adjust $\Omega^2$ ({\it i.e.} the
light beam intensity)
in order to get a quadrupole to dipole couplings ratio close to 1. The off-diagonal density matrix element holding the Berry Phase, decays to half its initial value. But adjusting the  interferometer parameters can  
compensate for this effect and keep the interference contrast close to its maximum. 

On the above example the role of both the hyperfine interaction, and the instability of the atomic excited states is clearly exhibited. This gives a clear illustration of  the atomic  internal structure contribution  to  the spin dynamics.  Although these  effects upon  Berry's cycles  can be accounted for by choosing appropriate values of the effective $\vB,\vE$ fields,  they lead to  severe experimental constraints. 
It looks possible to satisfy these constraints with $^{87}$Rb atoms but with light alkalis this is far from obvious \cite{bou3}. 
 On the other hand, the external degrees of freedom, {\it i.e.} the atomic nuclei coordinates have been supposed fixed physical quantities, an assumption which would obviously fail for ultra cold atoms belonging to a fermionic or bosonic quantum gas. In such a case, these coordinates become truly quantum variables which appear explicitly  in the density matrix of the system. The problem of Berry's phase generated by an adiabatic cycle of the coupled external fields is well beyond the scope of the present paper. 
 
{\bf Acknowledgments} 

The authors thank Dr. M. D. Plimmer for helpful discussions.
\vspace{-2mm}
 \appendix 
 \section*{ Appendix. Some useful formulas and  algebraic results for S=3 and S=4.}
\be
{\cal H}_{odd}(3,\lambda)=\left(
\begin{array}{cccc}
 \frac{3 \lambda }{2}+3 & \frac{\sqrt{15} \lambda }{2} & 0 & 0 \\
 \frac{\sqrt{15} \lambda }{2} & \frac{11 \lambda }{2}+1 & 3 \lambda  & 0 \\
 0 & 3 \lambda  & \frac{11 \lambda }{2}-1 & \frac{\sqrt{15} \lambda }{2} \\
 0 & 0 & \frac{\sqrt{15} \lambda }{2} & \frac{3 \lambda }{2}-3
\end{array}
\right)
\ee
\be
{\cal H}_{even}(3,\lambda)=
\left(
\begin{array}{ccc}
 4 \lambda +2 & \sqrt{\frac{15}{2}} \lambda  & 0 \\
 \sqrt{\frac{15}{2}} \lambda  & 6 \lambda  & \sqrt{\frac{15}{2}} \lambda  \\
 0 & \sqrt{\frac{15}{2}} \lambda  & 4 \lambda -2
\end{array}
\right)
\ee
\onecolumngrid
\be
  \det({\cal H}_{odd}(3,\lambda) -x \;\mathbf{l})=x^4-14 x^3 \lambda +49
x^2 \lambda ^2-10 x^2-36 x \lambda ^3+102 x \lambda -216 \lambda ^2+9 ,
  \ee
  
   \be
  \det({\cal H}_{even}(3,\lambda) -x \;\mathbf{l})=x^3-14 x^2 \lambda +49
x \lambda ^2-4 x-36 \lambda ^3+24 \lambda .
\ee
\be
{\cal H}_{even}(4,\lambda)=
\left(
\begin{array}{ccccc}
 2 \lambda +4 & \sqrt{7} \lambda  & 0 & 0 & 0 \\
 \sqrt{7} \lambda  & 8 \lambda +2 & 3 \sqrt{\frac{5}{2}} \lambda  & 0 & 0 \\
 0 & 3 \sqrt{\frac{5}{2}} \lambda  & 10 \lambda  & 3 \sqrt{\frac{5}{2}}
\lambda  & 0 \\
 0 & 0 & 3 \sqrt{\frac{5}{2}} \lambda  & 8 \lambda -2 & \sqrt{7} \lambda  \\
 0 & 0 & 0 & \sqrt{7} \lambda  & 2 \lambda -4
\end{array}
\right)
\ee
\be
{\cal H}_{odd}(4,\lambda)=
\left(
\begin{array}{cccc}
 \frac{11 \lambda }{2}+3 & \frac{3 \sqrt{7} \lambda }{2} & 0 & 0 \\
 \frac{3 \sqrt{7} \lambda }{2} & \frac{19 \lambda }{2}+1 & 5 \lambda  & 0 \\
 0 & 5 \lambda  & \frac{19 \lambda }{2}-1 & \frac{3 \sqrt{7} \lambda }{2} \\
 0 & 0 & \frac{3 \sqrt{7} \lambda }{2} & \frac{11 \lambda }{2}-3
\end{array}
\right)
 \ee
  \bea
  \det({\cal H}_{even}(4,\lambda) -x \;\mathbf{l}) &=&x^5-30 x^4 \lambda
+273 x^3 \lambda ^2-20 x^3-820 x^2 \lambda ^3+472 x^2 \lambda,
\eea
   \bea
 \det({\cal H}_{odd}(4,\lambda) -x \;\mathbf{l})=
x^4-30 x^3 \lambda +273 x^2 \lambda ^2-10 x^2-820 x \lambda ^3+
182 x \lambda +  576  \lambda ^4-712 \lambda ^2+9.
\eea
\twocolumngrid 

 \end{document}